%% file: main.tex
\documentclass[sigconf]{acmart} 
\usepackage{caption}
\usepackage{subcaption}
\usepackage{graphicx}
\usepackage{booktabs}

\usepackage{multirow}
\usepackage{longtable}

\usepackage[utf8]{inputenc}
\usepackage{microtype}

\usepackage{array,framed}
\usepackage{xspace,acmart-taps}
\usepackage{placeins}

\usepackage{xcolor}
\usepackage{colortbl}
\usepackage{collcell}
\usepackage{xstring}
\usepackage{textcomp}

\usepackage{hyperref}

\usepackage{rotating}
\usepackage{arydshln}
\usepackage{enumitem}
\usepackage{pgf}
\usepackage{calc}

\usepackage{wrapfig}
\usepackage{framed}

\usepackage{soul}
\usepackage{fancybox}

\newcommand{\hlc}[2][yellow]{{%
    \colorlet{foo}{#1}%
    \sethlcolor{foo}\hl{#2}}%
}

\definecolor{bg}{RGB}{255,255,255}
\definecolor{bgbord}{RGB}{235,235,235}
\definecolor{extitle}{HTML}{455a64}
\newenvironment{examplebox}[1]
  {
    
    \MakeFramed{\advance\hsize-\width\FrameRestore}
    \noindent\textbf{\textcolor{extitle}{#1}}\par\nobreak\smallskip 
    \setlength{\parskip}{4pt}
    \setlength{\parindent}{4pt}
  }
  {
    \endMakeFramed
  }

\newcommand{\AAAcolor}[1]{\textcolor{black!80}{#1}}
\newcommand{\promptcolor}[1]{\textcolor{gray!80}{#1}}
\newcommand{\outputcolor}[1]{\hlc[green!30]{#1}}

\newcommand{\minput}{\AAAcolor{\textsc{input}}\xspace}
\newcommand{\mprompt}{\promptcolor{\textsc{prompt}}\xspace}

\newcommand{\moutput}{\colorbox{green!30}{\textsc{output}}\xspace}

\definecolor{cone}{HTML}{757575}
\definecolor{ctwo}{HTML}{bdbdbd}
\definecolor{cthree}{HTML}{d32f2f}

\newcommand{\cone}{\hlc[cone!30]{\textsc{Know}}\xspace}

\newcommand{\ctwo}{\hlc[ctwo!30]{\textsc{Search}}\xspace}

\newcommand{\cthree}{\hlc[cthree!30]{\name}\xspace}

\newcolumntype{P}[1]{>{\arraybackslash}m{#1}}

\newcommand\BibTeX{B\textsc{ib}\TeX}
\definecolor{oran}{HTML}{aa8c2c}
\AtBeginDocument{%
  \providecommand\BibTeX{{%
    \normalfont B\kern-0.5em{\scshape i\kern-0.25em b}\kern-0.8em\TeX}}}

\copyrightyear{2024}
\acmYear{2024}
\setcopyright{rightsretained}
\acmConference[CHI '24]{Proceedings of the CHI Conference on Human Factors in Computing Systems}{May 11--16, 2024}{Honolulu, HI, USA}
\acmBooktitle{Proceedings of the CHI Conference on Human Factors in Computing Systems (CHI '24), May 11--16, 2024, Honolulu, HI, USA}
\acmDOI{10.1145/3613904.3642054}
\acmISBN{979-8-4007-0330-0/24/05}

\begin{document}
\newcommand{\icona}{{\includegraphics[height=1.2\fontcharht\font`\B]{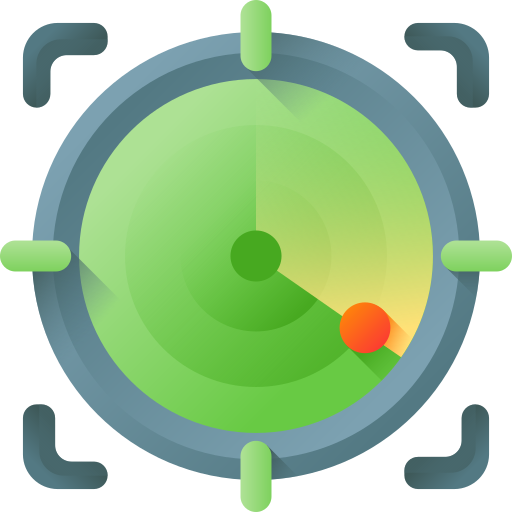}}\xspace}
\newcommand{\namecolor}{\textsc{Blip}}
\newcommand{\name}{\textsc{Blip}\xspace}
\newcommand{\blip}{\textbf{\fontfamily{ppl}\selectfont BLIP}\xspace}
\newcommand{\namewospace}{\textsc{Blip}}

\definecolor{labbg}{RGB}{246, 178, 49}

\newcommand{\ilabel}[1]{
    \begin{picture}(7,8) 
      \put(3,3.5){\color{labbg}\circle*{11}} 
      \put(3.5,4){\makebox(0,0){{\textcolor{black}{\scriptsize\bfseries\sffamily #1}}}} 
    \end{picture}
}

\title{\namecolor: Facilitating the Exploration of Undesirable Consequences of Digital Technologies}

\def\sector{domain\xspace}
\def\sectors{domains\xspace}
\def\Sector{Domain\xspace}
\def\Sectors{Domains\xspace}

\def\articlesTotal{16.3k\xspace} 
\def\articlesRelevant{4.8k\xspace} 

\def\paragraphsTotal{210k\xspace} 
\def\paragraphsRelevant{53k\xspace} 
\def\paragraphsConsequences{18k\xspace} 

\def\consequence{undesirable consequence\xspace} %
\def\consequences{undesirable consequences\xspace} %

\def\Consequence{Undesirable Consequence\xspace} %
\def\Consequences{Undesirable Consequences\xspace} %

\def\z{\phantom{0}}
\def\c{\phantom{,}}
\def\zc{\phantom{0,}}
\def\developers{developers\xspace}

\author{Rock Yuren Pang}
\affiliation{%
 \institution{Paul G. Allen School of Computer Science, \\ University of Washington}
 \streetaddress{185 E Stevens Way NE}
 \city{Seattle}
 \state{Washington}
 \country{USA}}

\author{Sebastin Santy}
\affiliation{%
 \institution{Paul G. Allen School of Computer Science, \\ University of Washington}
 \streetaddress{185 E Stevens Way NE}
 \city{Seattle}
 \state{Washington}
 \country{USA}}

\author{René Just}
\affiliation{%
 \institution{Paul G. Allen School of Computer Science, \\ University of Washington}
 \streetaddress{185 E Stevens Way NE}
 \city{Seattle}
 \state{Washington}
 \country{USA}}

\author{Katharina Reinecke}
\affiliation{%
 \institution{Paul G. Allen School of Computer Science, \\ University of Washington}
 \streetaddress{185 E Stevens Way NE}
 \city{Seattle}
 \state{Washington}
 \country{USA}}

\renewcommand{\shortauthors}{Pang et al.}
\newcommand{\draftonly}[1]{{\color{red}{#1}}}
\newcommand{\todo}[1]{\draftonly{#1}}
\definecolor{scolor}{HTML}{ff5722}
\newcommand{\temp}[1]{\todo{{\color{gray}[#1]$_{t}$}}}

\definecolor{revision}{HTML}{ff7043}
\newcommand{\rr}[1]{\textcolor{black}{#1}}

\definecolor{publisher}{HTML}{795548}
\newcommand{\publisher}[1]{\textcolor{publisher}{\emph{#1}}}

\begin{abstract}
Digital technologies have positively transformed society, but they have also led to undesirable consequences not anticipated at the time of design or development. 
We posit that insights into past undesirable consequences can help researchers and practitioners gain awareness and anticipate potential adverse effects. To test this assumption, we introduce 
\name, a system that extracts real-world \consequences of technology from online articles,
summarizes and categorizes them, and
presents them in an interactive, web-based interface.
In two user studies with 15 researchers in various computer science disciplines, we found that \name substantially increased the number and diversity of undesirable consequences they could list in comparison to relying on prior knowledge or searching online. Moreover, \name helped them identify \consequences relevant to their ongoing projects, made them aware of undesirable consequences they ``had never considered,” and inspired them to reflect on their own experiences with technology. 
\end{abstract}




\maketitle

\input{sections/01-introduction}
\input{sections/02-related_work}
\input{sections/03-methods}
\input{sections/04-study}
\input{sections/05-result}
\input{sections/05.5-new_study}

\input{sections/06-discussion}

\input{sections/09-future}

\begin{acks} 
\rr{We thank our participants and the anonymous reviewers for their valuable feedback. We also thank Sandy Kaplan, Alicia Gao, Kevin Feng, and Yadi Wang for their helpful suggestions. This work was funded by the National Science Foundation as part of the awards IIS-2006104, ER2-2315937, and SMA-2315937. 
}
\end{acks}


\bibliographystyle{ACM-Reference-Format}
\bibliography{main.bib}


\end{document}

%% file: sections/01-introduction.tex
\section{Introduction}

With the rise of digital technologies in our lives, society has not only experienced their benefits but also increasingly their undesirable consequences. Research and news headlines describe seemingly unavoidable side effects of digital technologies, from Instagram's adverse effects on adolescent girls' body images~\cite{kleemans2018picture} to Microsoft's chatbot Tay using racist language~\cite{wolf2017we}. Technological progress is commonly seen as a moral commitment that is ``legitimized no matter how dangerous''~\cite[p.325]{Parvin2020UnintendedBD}. While \consequences are sometimes described as accidental and minor \emph{``blips''}, researchers, journalists, and policymakers have suggested that many cases could have been avoided if technology developers were aware of similar issues and had taken cautious evaluation beforehand~\cite{Matthews2022EmbracingCV,bruckman2020have}. 

However, anticipating the various outcomes of technology is difficult~\cite{nanayakkara2020anticipatory}. 
In fact, recent work has found that computer science (CS) researchers at the forefront of developing new technologies are eager to proactively consider \consequences of their innovations, but lack well-formulated processes and tools to do so effectively~\cite{Do2023ThatsIB}. 
They reported that not having resources that provide a comprehensive understanding of ``common problems'' reduced their ability to anticipate \consequences~\cite[p.7]{Do2023ThatsIB}. 
Could insights into past undesirable consequences of technology help them gain awareness of potential future consequences?

\rr{We study this question by collecting a catalog of ``common problems,'' allowing CS researchers to explore past \consequences as reported in technology magazines and research papers.} Learning from prior incidents has proven to be useful in several settings, ranging from exploratory forecasting of technological advances~\cite{martino2003review}, improving software by studying a collection of previous defects~\cite{just2014defects4j}, to training future pilots using the aviation accidents database~\cite{jav508336}. Incorporating known and real-world case studies of ethical dilemmas into an undergraduate Human-Centered Computing class has also been shown to amplify students' engagement in ethical thinking~\cite{skirpanEthics2018}. What remains unknown is whether providing CS researchers with examples can have similar advantages, increasing their awareness of the various societal impacts of technology and supporting them in considering the potential consequences of their own projects. \rr{A challenge in the domain of \consequences is the lack of such resource across diverse CS subfields, and the unclear impact on CS researchers, who often lack the time for such in-depth consideration for diverse consequences~\cite{Do2023ThatsIB, srikumar2022advancing}.}

Hence, the goal of this paper is to explore whether providing CS researchers with a catalog of ``common problems'' would improve their awareness of \consequences. 
Our secondary goals are to find out how we can feasibly collect a self-updating catalog given that this information is currently scattered and the technology landscape is fast-moving, and how a system providing this service would be perceived and used by CS researchers.
We tackle these questions by designing, developing, and evaluating \name, a prototype system that collects and showcases \emph{a catalog of \consequences of digital technologies}. \name (1) automatically extracts real-world \consequences of technology from any given online article using natural language processing (NLP) techniques, (2) summarizes and categorizes them based on the aspect of life that they affect (such as health, equality, or politics), and (3) presents them in an interactive, web-based interface (see \autoref{fig:interface}). Users can use \name to view, sort, and save the currently 5.7k summaries of \consequences, or extract \consequences from additional articles. 

While considering \consequences is not yet a common practice among researchers, we designed \name to facilitate this process in the future.
We tested our assumption about \name's usefulness in two user studies. 
In the first study with nine CS researchers, we assessed \name's overall usefulness to consider 
\consequences in their broader field (e.g., social media) compared to two alternative approaches---relying on their prior knowledge and searching for \consequences online---and conducted in-depth interviews to further understand users' perceptions of \name and potential use cases. 
Our results show that \name enabled participants to add an average of 7.00 more \consequences beyond those they could list when relying on prior knowledge and searching online.  
Participants perceived \name as improving their ability to ``think outside the box'', made them aware of consequences that they ``had never considered,'' and was an essential way to collect \consequences ``because you can't just read a bunch of disconnected articles about this [and make sense of it].''

In our second study with six CS researchers, we followed up on these results, evaluating whether \name is useful and actionable in the context of specific projects that participants \rr{work on} across CS subdisciplines. All participants could find several \consequences relevant to their specific projects in less than 15 minutes, on average. Some of these were immediately actionable.
Overall, this paper contributes:

\begin{enumerate}
\item \rr{Empirical evidence that a catalog of \consequences supports CS researchers} in considering more, and more diverse, \consequences than if they rely on their prior knowledge or an online search (Study 1) and that it enables them to uncover potential adverse effects of their own projects (Study 2).  
\item An open-source, web-based system, \name\footnote{https://blip.labinthewild.org/}, that collects, summarizes, and categories \consequences. To develop \name, we designed an information distillation pipeline that leverages NLP techniques to efficiently establish a self-updating catalog of \consequences.
\end{enumerate}

%% file: sections/02-related_work.tex
\section{Related Work}

We use the term ``undesirable consequences'' to refer to negative consequences of digital technology that affect society~\cite{de2015unintended}. Oftentimes, undesirable consequences are unanticipated or even unintended~\cite{merton1936unanticipated}. We chose to work with ``undesirable consequences'' over the more prevalent ``unintended consequences'' of technology to emphasize that our primary concern is with exploring the adverse effects of technology.
Research in HCI and Science and Technology Studies (STS) has contributed a large body of work on observed negative effects of digital technology domains and products, including mobile phones~\cite{reyns2013unintended,Moser:2016}, the sharing economy~\cite{dillahunt2016does}, machine learning~\cite{cabitza2017unintendedCO,cabrera2019fairvis}, and social media~\cite{del2016spreading,starbird2019disinformation}. Researchers have also described various aspects of our lives that may be adversely affected by technology, such as its impacts on the environment~\cite{borning2020invisible}, health~\cite{harrison2007unintended,ash2007categorizing}, or privacy~\cite{acquisti2015privacy}. Moreover, researchers have increasingly investigated and brought to our attention differential effects of digital technology on certain population groups, such as on different gender~\cite{ekstrand2018all} and racial groups~\cite{schlesinger2018let, buolamwini2018gender}, low-income and underserved communities~\cite{dillahunt2016does}, or people in other countries~\cite{toyama2015geek, pang2023auditing, reinecke2011improving, santy-etal-2023-nlpositionality}.

\paragraph{\textbf{Discussions and interventions for addressing \consequences in research}}
With the increasing awareness of the potential adverse effects of technological innovations, the research community has started to engage in several efforts to prevent such incidents. For example, researchers have developed guidelines for ethical research and development~\cite{amershi2019guidelines,pairguidebook}, started dedicated conferences, such as FAccT, AIES, EAAMO, SIGCAS, and dedicated tracks (e.g., Critical Computing@CHI), led workshops~\cite{conseuqences} and ethical committees~\cite{ethics_committee,chairs_2021}, as well as called for changes in institutional structures~\cite{bernstein2021esr, pang2023case}, critical education~\cite{ko2020time}, and in how we address undesirable consequences of digital technologies~\cite{Matthews2022EmbracingCV,bruckman2020have}. A key concrete step was the inclusion of broad ethics or impact statements in major conferences, such as IUI~\cite{IUI2022}, NeurIPS~\cite{NeurIps2021}, ACL~\cite{ACL2022}. Nanayakkara et al.~\cite{nanayakkara2021unpacking} found that such statements diversified thinking about how ML research could potentially impact society, though they tended to focus on positive impact~\cite{ashurst2022}.  Additionally, there are calls for researchers within different computing communities to accurately report the design considerations of their datasets~\cite{gebru2021datasheets,bender2018data}, models~\cite{mitchell2019model,rogers2021changing}, and tasks~\cite{Lindley2017ImplicationsFA,mohammad2021ethics} as well as evaluate and de-bias their products~\cite{amershi2019guidelines, pairguidebook, Bellamy2019AIF3, Wexler2020TheWT}. 

\paragraph{\textbf{Methods for forecasting and anticipating \consequences}}
Researchers have designed tools to help identify and contemplate social values of different stakeholders, such as the Envisioning Cards~\cite{Friedman2012TheEC}, Tarot Cards of Tech~\cite{TarotCardofTech}, and Value Cards~\cite{Shen2021ValueCA} (see also~\cite{chivukula2021surveying} for a detailed overview). The value-sensitive design approach has also contributed broad guidelines for researchers seeking to account for human values in a principled and systematic manner throughout the design process~\cite{Friedman2013ValueSD, Zhu2018ValueSensitiveAD}. The Future Ripples method~\cite{Epp2022ReinventingTW}, inspired by the Futures Wheel foresight method in education~\cite{Glenn1972FuturizingTV}, allows collaborative brainstorming on the impact of innovation through workshop activities. 

However, some of these approaches and methods have been challenged for not sufficiently supporting practitioners and the reality of the product-development process~\cite{gray2019ethical}. Using these methods requires prior knowledge on the topic at hand, which may not always be the case for novice users. They also require developers to deliberate in a team, sometimes with external experts, simultaneously and collectively where envisioned consequences can vary depending on the team's diversity and backgrounds. In fact, an interview with 20 CS researchers found that none of the participants are actively using these tools in practice~\cite{Do2023ThatsIB}. 

An alternative approach for anticipating undesirable effects of technology is \rr{learning from past incidents}~\cite{mcgregor2021preventing, yampolskiy2019predicting}. While perfectly predicting the future may be impossible, researchers have developed various methods to estimate what may happen from such past experiences. For example, the Delphi forecasting method~\cite{Weaver1971TheDF, Shin1998UsingDF} has been used in a wide variety of domains such as predicting air travel~\cite{English1976THEPO} and designing educational technology~\cite{Nworie2011UsingTD}. Another forecasting method is the case study method, which collects people’s thoughts on and experiences with past technology developments in an organization~\cite{Cheng2008AFM}. \rr{One issue} for the widespread use of these methods is that they usually rely on experts to collect and interpret historical data, making it difficult to scale and frequently use them.  

In another attempt to anticipate \consequences, prior work has developed a forum to collect news articles about technologies~\cite{neumann2008risks} and an AI incident database specifically for the effects of AI technology on society~\cite{mcgregor2021preventing}. One of the motivations for this database was that ``the artificial intelligence system community has no formal systems whereby practitioners can discover and learn from the mistakes of the past''~\cite[p.1]{mcgregor2021preventing}. 
However, the AI incident database necessitates the crowd to manually browse and enter incidences, using a leaderboard to incentivize contributions from volunteers. 
To the best of our knowledge, no previous system exists that automatically and systematically catalogs, summarizes, and categorizes \consequences for a variety of technology domains. None of these approaches have been formally evaluated to show their usefulness for anticipating \consequences.

\rr{In short, many prior tools prompt users to reflect on high-level ethical questions, which requires users to have prior knowledge without easy access to updated real-world examples. This paper explores the value of providing researchers with concrete examples that can ground their ethical considerations in practice. }

\vspace{-0.25cm}
\paragraph{\textbf{Supporting ideation of potential \consequences}}
\name was also inspired by work on creativity support tools, which showed that a collection of diverse examples can support ideation~\cite{siangliulue15:providing, pang2023anticipating}. For example, sampling diverse inspirational examples (and providing a visual overview of the ideas) has been found to improve people's brainstorming activity~\cite{siangliulue16:ideahound}. Similar work on cognition and creativity support confirmed that examples inspire and unveil new and diverse ideas~\cite{Ngoon2021ShwnAC, zhang2021method, Kang2018ParagonAO, Ngoon2018InteractiveGT, DiFede2022TheIM}. To organize these examples, prior work leveraged categories of certain topics and characteristics, which are essential to human cognition~\cite{Rosch1976BasicOI, Ward1994StructuredIT}. For example, IdeaRelate facilitates the exploration of COVID-related examples by tagging them into different topics, helping users to include more perspectives in their own idea generation~\cite{idearelate}.  Recent work attempted to incorporate \rr{language models} to help ideate potential harms ~\cite{park2022social, buccinca2023aha}. In particular, AngleKindling used few-shot LLM prompts to find potential controversies and negative outcomes from press releases to help journalists generate story ideas~\cite{petridis2023anglekindling}. However, zero-shot and few-shot approaches to generate consequences had resulted in rather generic results~\cite{petridis2023anglekindling}. In this work, we aid the inherently creative process of reflecting on past and possible future adverse effects by providing a catalog of \consequences, supplemented with information on the diverse aspects of life that they have affected. 
Instead of relying on ideas generated entirely from language models, we extract relevant information directly from a wide range of online articles, provide access to the original content, and update our collection \rr{every week}. 

%% file: sections/03-methods.tex
\section{\name} 
\name was developed to explore the value of providing \rr{researchers} with a catalog of past undesirable consequences of technology. Showing this catalog aims to address gaps where developers often overlook potential adverse impacts~\cite{Do2023ThatsIB}. \name's open-source code is available at \url{https://github.com/rrrrrrockpang/blip}, and its interface is currently deployed at \url{https://blip.labinthewild.org}. 

\vspace{-0.4cm}
\subsection{Design Choices and Rationale}
\label{sec:design_choice}

We followed a \rr{user-centered design} process in which we iteratively sought user feedback on several prototype interfaces before arriving at the present implementation. The design choices were also informed by prior literature as follows: 

\textbf{Everything in one place:} Combining information across the Internet is difficult, which is why several systems address the need to collect information in one place (e.g., Pinterest, Fuse~\cite{kuznetsov2022fuse}). \name is therefore designed as a web-based system that allows viewing and organizing \consequences across various technology domains in one place.

\textbf{Automatically collecting information:} 
Prior work showed that \developers desire a collection of past technology incidences, but that they lack the time to invest in collecting resources themselves~\cite{Do2023ThatsIB}. Moreover, manual curation takes time and requires motivating users to contribute this data, which can be difficult and result in a limited number of \consequences examples. 
To support the scale needed to achieve a fairly comprehensive and updated collection, we developed an approach for \emph{automatically} retrieving \consequences from a set of trusted online articles that regularly report on them. \name currently retrieves articles from reputed outlets that often report on new technologies, such as MIT Technology Review, TechCrunch, The Verge, and WIRED. This list can be easily expanded.

\textbf{Summarizing \consequences}: 
To help users effectively process online information, prior systems have summarized complex content in other contexts such as for reading papers~\cite{august2022paper}, reviewing academic literature~\cite{kang2022threddy} and conversing online~\cite{zhang2017wikum}. Similarly, we present users with a summary of any \consequences in an article. 
Our decision was further informed by a design and feedback session with three CS \rr{researchers}, in which we presented early mock-ups and discussed potential changes. \rr{Participants noted that the original articles were too long for a quick overview}, though all wanted to retain the possibility to access them. \rr{Participants noted} that seeing entire paragraphs prevented them from quickly understanding what the societal implication is. Instead, we decided to provide a summary specific to the \consequences in an article. 

\textbf{Categorizing \consequences}: 
Prior work has examined the benefits of category structure in human cognition for sensemaking and creativity~\cite{Ward1994StructuredIT}. More recent systems have leveraged different ``categories'' to organize mass information~\cite{kuznetsov2022fuse} and generate creative ideas~\cite{idearelate, siangliulue16:ideahound, Ngoon2018InteractiveGT}. 
In addition, our decision was confirmed in the same preliminary study above, in which participants reported that presenting \consequences without any organization was overwhelming and time-consuming. To address this issue, we designed \name to categorize \consequences into different aspects of life that they affect, such as politics and equality, which we adapted from the Tarot Cards of Tech~\cite{TarotCardofTech}. \name visually signals these categories with different colors, which may enhance users' understanding of the diverse range of \consequences that can occur in the real world.

\textbf{Bookmarking articles}: In early feedback on our prototype, we also repeatedly received the feedback that users wanted to return to a specific article or save it as a collection of \consequences particularly relevant to their project. \name therefore allows bookmarking articles in a sidebar using cookies.

 \subsection{User Interface Usage Scenario}
 \label{sec:usage_scenario}

 \begin{figure*}
    \centering
    \includegraphics[width=0.95\textwidth]{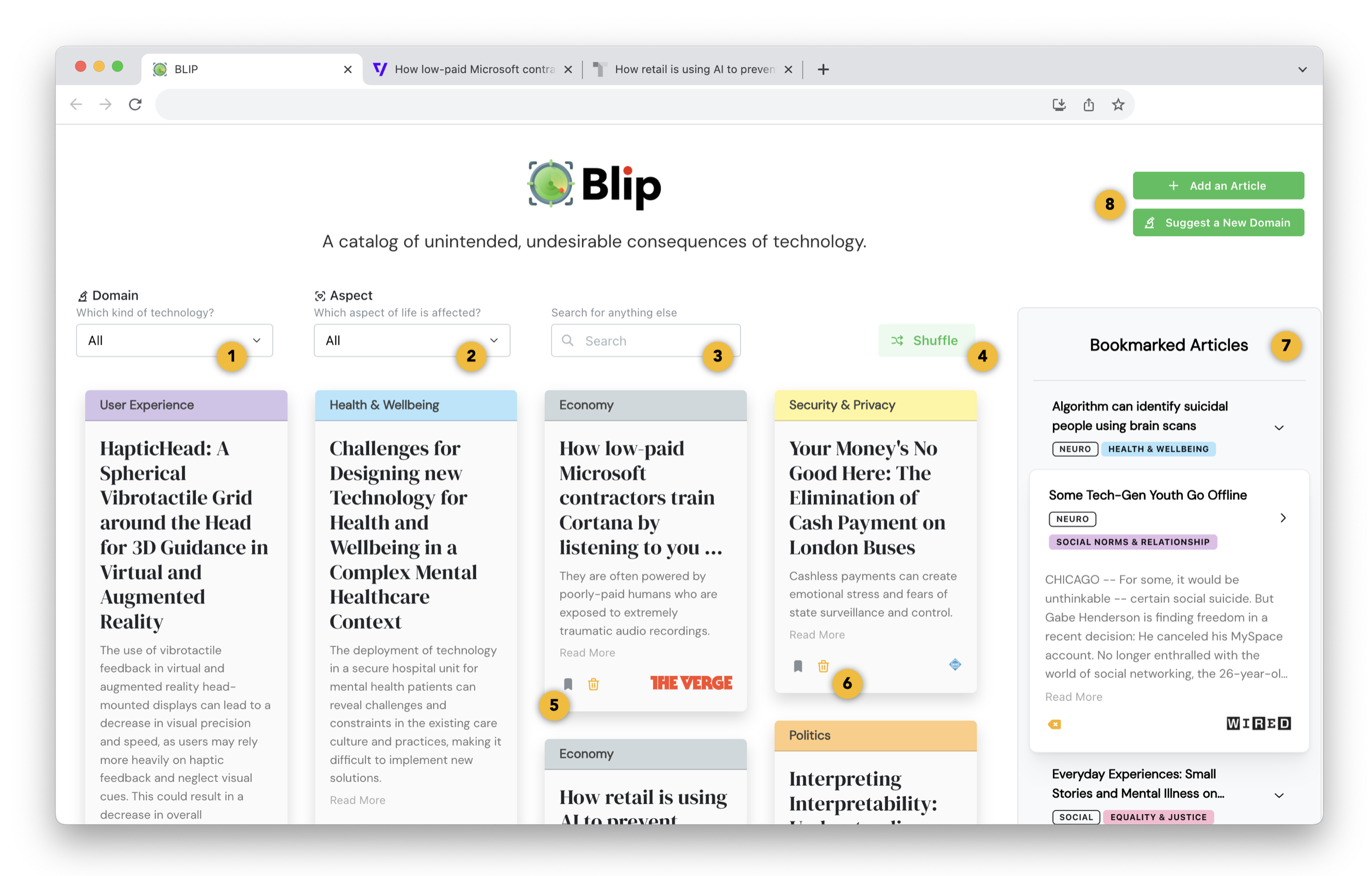}
    \vspace{-0.5cm}
    \caption{\name's main user interface. Users can view summaries of \consequences and filter them by technology domains, aspects of life they affect, or search keywords. }
    \Description{Shows \name's user interface. It is a card-based interface on which users can select between three technology domains to review summaries of undesirable consequences from online articles. They are also provided with other features such as searching with terms, selecting aspects of consequences and a shuffle button to allow for getting new ideas.}
    \label{fig:interface}
    \vspace{-0.5cm}
\end{figure*}

 In this section, we illustrate how users can interact with \name's user interface to explore undesirable consequences.  At a high level, the main interface (\autoref{fig:interface}) allows users to (1) browse through diverse examples of \consequences for different technologies, (2) understand and access the source articles, (3) filter and search \consequences, and (4) bookmark articles, e.g. if they wanted to read the article later or create a subset of \consequences for later consideration. 
 
 \name displays different \consequences on cards in a scrollable interface (see~\autoref{fig:interface}). As the user scrolls down, \rr{new cards appear automatically.} Each card includes a header that displays the aspect of life it affects in a distinct color to promote visual organization. The card content includes the summarized undesirable consequence along with the article title and source, as well as two buttons that let users bookmark or delete an article from the view. Clicking on the article title opens a new browser tab that shows the original article. Bookmarked cards appear on a history sidebar \ilabel{7}.
 By default, \name shows all cards in random order, but users can filter the cards by technology domain \ilabel{1} and/or by the aspect of life  \ilabel{2}. They can also search for specific terms within the summary, such as ``mental health'' or ``misinformation'' \ilabel{3}. The shuffle button at the top allows users to shuffle cards in the collection view to encounter new ideas \ilabel{4}.
 Users can save a card \ilabel{5} to their bookmark \ilabel{7}. When users think that they have already known a consequence in a card, they can remove that from their view \ilabel{6}. Users can review their collection of articles at \ilabel{7} to gain awareness of the consequences discussed online.
 Users can also import an article via an article URL in \ilabel{8} as described in Section~\ref{extension}.

\vspace{-0.2cm}
\begin{figure*}[ht]
   \includegraphics[width=0.9\linewidth]{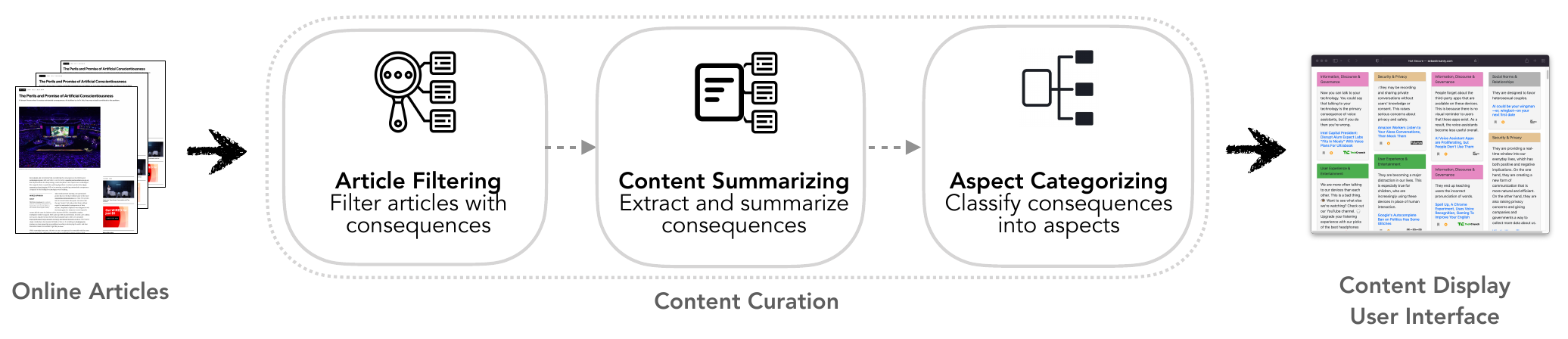}
   \vspace{-0.5cm}
   \caption{An overview of \name's content curation pipeline. Given online article sources, \name filters out those that discuss \consequences, extracts and summarizes the consequence, and categorizes it into different aspects of life that it affects, such as the environment, equality, or politics. The \consequences are then displayed in an interactive, web-based user interface.  
   }
   \label{fig:overview}
 \end{figure*}

\subsection{Content Curation Pipeline}

As shown in \autoref{fig:overview}, \name automatically filters relevant articles describing undesirable consequences of given input articles in a technology domain, (Section \ref{sec:filtering}), extracts and summarizes these consequences (Section \ref{sec:summarizing}), categorizes them into different aspects of life and society that they affect (Section \ref{sec:categorization}), and finally displays them in an interactive interface in \autoref{fig:interface}.
To achieve these steps, \name uses GPT-3.5~\cite{brown2020language} due to its versatility and high-quality outputs. GPT-3.5~\cite{brown2020language} is noted for its ability to classify with higher accuracy than supervised approaches with no or few training instances. We used the gpt-3.5-turbo model, a pre-trained language model that can solve NLP tasks with instructions. The model can be accessed via its OpenAI API~\cite{openai}. To show the model how to perform a given task, it has to be given instructions along with examples. Such `zero-shot' methods benefit our case~\cite{Wei2021FinetunedLM} since annotating articles (for supervised approaches) is expensive because of the length of articles and relatively infrequent descriptions of \consequences. Hereon, we use the terms \minput to denote the input text, \mprompt to denote the natural language instruction, and \moutput to denote the output by the model.

\subsubsection{Article Filtering}
\label{sec:filtering}
 Given a large volume of input articles, filtering relevant articles that contain  \consequences is our first step. 
 \name performs filtering in two steps based on: (1) the title and (2) the content. This hybrid filtering method aims to include more relevant \consequences from articles and reduces the cost of computation from requesting the OpenAI API.

 \paragraph{Filtering by Title} 
 First, \name determines whether an article mentions \consequences based on the title information. For example, the title ``Social media is polluting society. Moderation alone won’t fix the problem'' is very likely to discuss such consequences. In contrast, titles that announce product launches, analyze products, or discuss corporate leadership rarely contain relevant consequences (e.g., ``Improbable teams with Google, opens SpatialOS alpha for virtual world development''~\cite{lunden_2016}). 

 To filter articles by title for those that mention \consequences, \name employs a RoBERTA-based~\cite{liu05:quantifying} supervised binary classifier that outputs whether an article is relevant or not. To develop this title classifier, we annotated a dataset of 1,500 random online article titles for whether they are likely to contain an undesirable consequence or not. 
 Two authors individually annotated all the article titles with a binary label ``relevant'' or ``irrelevant.'' The initial inter-rater reliability was 92.17\%. The two authors then discussed the inconsistent titles until agreement was achieved. When the two authors were unsure about the relevance during the deliberation phase, we included the titles and resorted to filtering by content to reduce ambiguity.
 Because obtaining diverse positive examples is more difficult than getting negative examples, we leveraged the AI incidents database~\cite{mcgregor2021preventing} to find articles that discuss undesirable consequences. The title classifier that was fine-tuned on this dataset achieved F1=86.63\% on a 4:1 train-test split. The performance of our title classifier is significantly higher than that of classifiers in comparable prior work (e.g., when detecting propaganda in news articles, where an F1 score of 60.98\% was reported~\cite{da2019fine}).

\vspace{-0.2cm}
\paragraph{Filtering by Content} 

 \name additionally filters the article content using the prompting approach~\cite{brown2020language}. 
 More precisely, \name uses the following \mprompt: ``Does the article above discuss unintended or undesirable consequences on society of <domain>? Answer Yes or No.'' An example \minput, \mprompt, and \moutput looks as follows:

\aptLtoX[graphic=no,type=html]{ 
\begin{framed}{\textbf{An Example of Filtering by Content} \\
\hphantom{000 0  0 0 0 }\\
\publisher{\small The Nauseating Disappointment of Oculus Rift (MIT Tech Review) ~\cite{metz1_2020}}}\\
\hphantom{000 0  0 0 0 }\\
	\texttt{\small
\AAAcolor{
    Admittedly, I was using a \$599 Oculus Rift virtual-reality headset. It was a lot of fun, though I looked like a complete idiot sitting with a clunky black gadget on my face. I also got a more in-depth look at simulator sickness—feelings of nausea, dizziness, and eye strain that some people get when using VR—and what it means for the future of this technology ... [continued]} \\
\\
    \promptcolor{Does the article above discuss undesirable consequences of virtual reality on society? Answer Yes or No.}\\ 
\\
{\colorbox{green!30}{Yes}}
}
\end{framed}
 }{ \begin{examplebox}{\textbf{An Example of Filtering by Content} \\
\publisher{\small The Nauseating Disappointment of Oculus Rift (MIT Tech Review) ~\cite{metz1_2020}}}
	\texttt{\small
\AAAcolor{
    Admittedly, I was using a \$599 Oculus Rift virtual-reality headset. It was a lot of fun, though I looked like a complete idiot sitting with a clunky black gadget on my face. I also got a more in-depth look at simulator sickness—feelings of nausea, dizziness, and eye strain that some people get when using VR—and what it means for the future of this technology ... [continued]} \\
    \promptcolor{Does the article above discuss undesirable consequences of virtual reality on society? Answer Yes or No.}\\ \outputcolor{Yes}
}
\end{examplebox}
 } 

\subsubsection{Content Summarizing}
\label{sec:summarizing}

Given the filtered set of articles, \name automatically summarizes \consequences. For example, consider the following paragraph from an article in the \textit{MIT Technology Review}~\cite{ryan-mosley_2021}: 

\begin{quote}
\emph{
    [...] social media was just making it worse. The prejudice Lise experienced—colorism—has a long history, driven by European ideals of beauty that associate lighter skin with purity and wealth, darker tones with sin and poverty. [...] And today, thanks to the prevalence of selfies and face filters, digital colorism has spread. With Snapchat, Instagram, TikTok, and Facebook a part of billions of people’s everyday lives, many of us find that people see far more pictures of us than ever before. [...]}
\end{quote}

While providing much detail and nuance, reading the entire article is time-consuming. We found that articles in our dataset included an average of 713.2 words (min=103; max=5401; SD=501.6). In our experience, we found it challenging to skim and locate phrases within paragraphs that describe undesirable consequences without getting caught up in the specifics. Therefore, listing various undesirable consequences discussed in articles was impractical at best because irrelevant details in the articles distracted from higher-level issues. Instead, a shorter summary of the discussed consequence helps users grasp the overall issue.
\name employs GPT-3.5 for abstractive summarization, which paraphrases the undesirable consequences discussed in the article and generates relatively short summaries. We leveraged LLMs given that prior work found LLM summaries to be on par with human written summaries~\cite{zhang2023benchmarking}. In contrast to extractive summarization, which extracts and concatenates excerpts from an article into a summary, abstractive summarization can generate summaries of a specified length and include multiple consequences, making it a better fit for our purpose.  
The \mprompt for this task is: ``To summarize in a short paragraph, the main undesirable consequence of <domain> being discussed here is''. An example \minput, \mprompt, and \moutput is as follows:

\aptLtoX[graphic=no,type=html]{ 
\begin{framed}{\textbf{An Example of Content Summarizing} \\
\hphantom{000 0  0 0 0 }\\\\
\publisher{\small How digital beauty filters perpetuate colorism (MIT Technology Review) ~\cite{ryan-mosley_2021}}}\\
\hphantom{000 0  0 0 0 }\\
\small
	\texttt{\small
\AAAcolor{When Lise was a young teenager in Georgia, her classmates bullied her relentlessly. She had moved with her family from Haiti a few years earlier, and she didn’t fit in with the other students. They teased her about her accent, claimed she “smelled weird,” and criticized the food she ate.  But most often they would attack her with remarks about her dark complexion ... \rr{As she struggled to cope with the abuse, Lise convinced herself that the darkness of her skin was to blame. And social media platforms and the visual culture of the internet suggested the same thing} ... [continued]} \\
\\
    \promptcolor{To summarize in a short paragraph, the main undesirable consequence of social media being discussed here: } {\colorbox{green!30}{It can lead to the reinforcement of colorism. Colorism is defined as prejudice against people with darker complexions, and it can have harmful effects on people's mental and physical health. Social media platforms like Instagram have filters that can lighten users' skin tone, which can perpetuate the idea that lighter skin is more desirable. In addition, recommendation algorithms on these platforms often favor content featuring people with lighter skin, which can reinforce users' biases and lead to the marginalization of people with darker complexions.} }
}
\end{framed}
 }{ \begin{examplebox}{\textbf{An Example of Content Summarizing} \\
\publisher{\small How digital beauty filters perpetuate colorism (MIT Technology Review) ~\cite{ryan-mosley_2021}}}
\small
	\texttt{\small
\AAAcolor{When Lise was a young teenager in Georgia, her classmates bullied her relentlessly. She had moved with her family from Haiti a few years earlier, and she didn’t fit in with the other students. They teased her about her accent, claimed she “smelled weird,” and criticized the food she ate.  But most often they would attack her with remarks about her dark complexion ... \rr{As she struggled to cope with the abuse, Lise convinced herself that the darkness of her skin was to blame. And social media platforms and the visual culture of the internet suggested the same thing} ... [continued]} \\
    \promptcolor{To summarize in a short paragraph, the main undesirable consequence of social media being discussed here: } \outputcolor{It can lead to the reinforcement of colorism. Colorism is defined as prejudice against people with darker complexions, and it can have harmful effects on people's mental and physical health. Social media platforms like Instagram have filters that can lighten users' skin tone, which can perpetuate the idea that lighter skin is more desirable. In addition, recommendation algorithms on these platforms often favor content featuring people with lighter skin, which can reinforce users' biases and lead to the marginalization of people with darker complexions.} 
}
\end{examplebox}
 } 


\subsubsection{Aspect Categorizing}
\label{sec:categorization}

\name assigns each undesirable consequence summary to one of 10 aspects of life, from health \& well-being to politics (see Supplementary Materials for a list of aspects and examples). This categorization allows for narrowing down the set of summarized  \consequences and can emphasize the variety of impacts technology can have on society. 

To develop the list of aspects, we built on the list of 21 aspects of life that are relevant to societal implications from the Tarot Cards of Tech project~\cite{TarotCardofTech}. Assigning 150 randomly chosen articles discussing \consequences to these 21 aspects of life, we iteratively merged and renamed the aspects to fit our data (see Supplementary Materials for details). The resulting 10 aspects of life broadly represent various categories that \consequences commonly fall into and are used in \name to support users in learning and brainstorming. We incorporated the list in \name such that it can be extended with additional aspects or replaced with a new list. 

\name uses the prompting approach of GPT-3.5 for aspect categorization. The \texttt{prompt} we use for this task is: ``Which aspect of life does the following consequence affect?'' An example \minput, \mprompt, and \moutput looks as follows:\\

\aptLtoX[graphic=no,type=html]{ 
\begin{framed}{\textbf{An Example of Aspect Categorizing} \\
\hphantom{000 0  0 0 0 }\\\\
\publisher{\small AI voice actors sound more human than ever—and they’re ready to hire (MIT Technology Review) ~\cite{hao_2021}}}
\hphantom{000 0  0 0 0 }\\
\small
\texttt{\promptcolor{List of possible aspects: Economy, Environment \& Sustainability, Equality \& Justice, Information \& Discourse, Health \& Well-being, Politics, Power, Security \& Privacy, User Experience \& Entertainment, Social Norms \& Relationships\\
Which aspect of life does the following consequence affect?\\
\\
Title: 
\AAAcolor{AI voice actors sound more human than ever—and they're ready to hire}\\
\\
Summary: 
\AAAcolor{People are losing their jobs. The technology is becoming so realistic that many people can't tell the difference.}\\
\\
Aspect:
}{\colorbox{green!30}{Economy}}
}
\end{framed}
 }{ \begin{examplebox}{\textbf{An Example of Aspect Categorizing} \\
\publisher{\small AI voice actors sound more human than ever—and they’re ready to hire (MIT Technology Review) ~\cite{hao_2021}}}
\small
\texttt{\promptcolor{List of possible aspects: Economy, Environment \& Sustainability, Equality \& Justice, Information \& Discourse, Health \& Well-being, Politics, Power, Security \& Privacy, User Experience \& Entertainment, Social Norms \& Relationships\\
Which aspect of life does the following consequence affect?\\
Title: 
\AAAcolor{AI voice actors sound more human than ever—and they're ready to hire}\\
Summary: 
\AAAcolor{People are losing their jobs. The technology is becoming so realistic that many people can't tell the difference.}\\
Aspect:
}\outputcolor{Economy}
}
\end{examplebox}
 } 
\subsubsection{Implementation Details and Costs}
\label{extension}
\name includes a frontend interface implemented in the React JavaScript library and a server using the FastAPI Python framework. The server uses Selenium~\cite{selenium} and Beautifulsoup~\cite{bs4} to extract article URLs based on input keywords and the newspaper3k API~\cite{newspaper3k} to obtain the article content. We used a combination of the sentence-transformers model in the huggingface library and the FAISS library to enable the quick search functions for similar articles to a search keyword~\cite{johnson2019billion}. In our main system architecture, we initially used the GPT-3 API and text-davinci-002 model~\cite{openai}, which was released on June 11, 2020. The cost of using the GPT-3 API was \$0.06 for ~750 words at the time of implementation. Since then, new variants of GPT models were made available, including GPT-3.5 and GPT-4, which were introduced after our first study. The language models that \name uses can be changed as more powerful versions come out. We also added an option to run the pipeline using open-source language models, Llama2. We re-ran our content curation pipeline on the three domains using GPT-3.5 on August 15, 2023.

\vspace{-0.2cm}
\subsection{Technical Evaluation}
\label{sec:case_study}

We evaluated the pipeline described above on three technology domains: social media (SM), virtual reality (VR), and voice assistants (VA). We chose these three domains as the initial content for \name because they represent diverse digital technologies that have been deployed and used for different amounts of time. Social media is a technology with widely-explored  consequences on economic, political, and social spheres (e.g., polarization~\cite{tucker2018social} and depression~\cite{cunningham2021social}). Voice assistants are comparatively new but are now an integral part of many people's lives, with consequences including privacy violations~\cite{natatsuka2019poster} and harmful content~\cite{iovine_2021}. Virtual reality is still newer and has not yet become mainstream. 

\paragraph{Retrieving Online Articles}
We searched for articles in these domains using the keywords below from the sources in \autoref{tab:title-classifier}. The search led to a total of 42,405 articles, published between 1997-2023.

\paragraph{Article Filtering}
Applying the title classifier to our dataset of online articles resulted in a total of 26,628 articles as shown in Table \ref{tab:title-classifier}.  
Filtering by content retained 2.6k articles in our dataset that discuss \consequences of SM, VA, and VR. This process filters out articles with ambiguous titles related to \consequences. For example, an article titled ``Advertisers Employ Social Media'' was predicted as relevant by title filtering. However, the article discusses companies that use social media for advertisement with no clear \consequences. The content-based classifier achieved an accuracy of 89.24\% (F1=89.83\%), which is a 3\% increase compared to the title classifier.

\input{files/tab_filtering.tex}

\begin{itemize}[leftmargin=0.15in] 
    \label{search_list}
    \item \textbf{Social Media}: \texttt{social media}\footnote{We avoided product-specific search terms to avoid capturing generic articles that included text, such as ``Share this link on Facebook and Twitter.''}
    \item \textbf{Voice Assistants}: \texttt{voice assistant, chatbot, home assistant, AI assistant, speech recognition, voice recognition, smart assistant, personal assistant}
    \item \textbf{Virtual Reality}: \texttt{virtual reality, mixed reality, augmented reality, metaverse}
\end{itemize}

\paragraph{Content Summarizing}

To evaluate the accuracy of the generated summaries in describing the \consequences from the articles, we randomly chose 50 articles from the filtered set. 
One author read each article and graded the corresponding summary as either accurate or inaccurate. A second author then confirmed the decision.  
The summary was considered accurate if it truthfully \rr{translated} the desired content (\consequences in our case). Sources of inaccuracy included (1) introducing facts absent in the original text (also known as model hallucinations~\cite{Maynez2020OnFA}), (2) failing to summarize \consequences because the articles never discuss \consequences, (3) producing a \rr{non-sensical summary} (a phenomenon known as text degeneration~\cite{holtzman2019curious}), or (4) generating oversimplified summary with insufficient context (requiring decontextualization~\cite{choi2021decontextualization}).
For example, the summary ``They [VA] will probably make us all look like idiots.'' is inaccurate because it is oversimplified (does not contain enough context to stand-alone). We found that 84\% (42/50) were rated as accurate, suggesting that summarizations are largely reliable. 
For example, one wrong summarization of an article on the Interpreter Mode of Google Translator ~\cite{Chokkattu_2019} is ``we will be often talking to our devices than each other. This is a bad thing.'' The article mentions potential mistranslation for people with thick accents but does not explicitly mention overuse. A future improvement could be allowing user feedback to enhance the potentially incorrect summaries.

\paragraph{Aspect Categorizing}

We evaluated \name's 10-way classification using the pilot 150 articles assigned to 10 different aspects of life. \name's classification achieved an accuracy of 38\% (F1=36\%), which is not ideal but expected as the performance is comparable to other multi-class classification tasks~\cite{rashkin2018modeling} (even including those with fewer categories). Achieving higher accuracy with the zero-shot approach is difficult. In our case, the fairly complex categories (e.g., Health \& Well-Being) and their potential for overlap lower the performance; misclassifications in our dataset are rarely egregious but instead happen when multiple categories could be a potential fit.

 \subsection{Growing \name's Content}
 \label{extension}
 While our technical evaluation demonstrates the feasibility of adapting NLP techniques to extract \consequences for three technology domains, \name includes two ways for adding more \consequences and additional technology domains. 
 
 First, users can click on the "Import an article" button in \name's user interface (\autoref{fig:interface}-\ilabel{8}) to add a single article via URL or PDF. \name then runs the article through its extraction pipeline and shows a card with the summary, link, and aspect category as output. Users can assign the card to an existing technology domain or propose a new one (e.g., robotics). Added articles, cards, and technology domains are stored in a temporary database and only added after approval to avoid potential misuse and retain control over the number of technology domains that are being added. 

 A second option is to use \name's bulk-import functionality, which is currently only available to developers to control the costs of accessing the OpenAI API. This functionality allows inputting a keyword (e.g., social media) or adding several URLs or an entire spreadsheet data file with several articles at once, which \name then runs through the extraction process that can take several hours. As we described in Section \ref{sec:case_study}, we have previously obtained online articles from a set of trusted online technology magazines by searching for, and downloading, those that discuss specific technology domains. We plan to continue using this approach for adding new domains (as we did to add more fields in our Study 2).

In addition to these two approaches for manually adding \consequences, the system automatically checks for new articles in the three domains (and on the four sites listed in Table~\ref{tab:title-classifier}) on a weekly basis and adds the discussed \consequences. The update frequency is flexible; we decided on weekly updates because there are usually only 3-5 new articles every week. 

%% file: files/tab_filtering.tex
\aptLtoX[graphic=no,type=html]{ \definecolor{smvavrcolor}{HTML}{9e9e9e}
\newcommand{\smvavr}[3]{\textcolor{smvavrcolor}{#1$\cdot$#2$\cdot$#3}}
\begin{table*}
\small
    \caption{List of online sources used for retrieving articles on the three technology domains in our technical evaluation: Social Media (SM), Voice Assistants (VA), and Virtual Reality (VR). The table also shows the percentage and total number of relevant articles that contain \consequences after each filtering step.}
    \label{tab:title-classifier}
    \vspace{-0.5cm}
    \begin{center}
        \begin{tabular}{ p{3.2cm} p{1.9cm} >{\raggedright\arraybackslash}p{2.4cm} >{\raggedright\arraybackslash}p{2.1cm}>{\raggedright\arraybackslash}p{2.1cm}}
        \toprule
        \textbf{News Source} & \textbf{Year Range} & \textbf{Retrieved Articles \newline {\footnotesize \smvavr{SM}{VA}{VR}}} & \textbf{Title Filter \newline {\footnotesize \smvavr{SM}{VA}{VR}}} & \textbf{Content Filter \newline {\footnotesize \smvavr{SM}{VA}{VR}}} 
        \\
        \midrule
        \href{https://www.technologyreview.com/}{MIT Technology Review} & 1997-2022 
        & 3433 \newline {\footnotesize \smvavr{1686}{957}{790}}  
        & 1957 \hfill {(57\%)} \newline {\footnotesize \smvavr{1082}{563}{312}}   
        & 519 \hfill {(15\%)} \newline {\footnotesize \smvavr{349}{116}{54}}  
        \\
        \hdashline\noalign{\vskip 0.1cm}
        
        \href{https://techcrunch.com/}{TechCrunch} & 2005-2022 
        & 3975 \newline {\footnotesize \smvavr{748}{1502}{1725}}
        & 1330 \hfill {(33\%)} \newline {\footnotesize  \smvavr{337}{538}{455}} 
        & 390 \hfill {(10\%)} \newline {\footnotesize \smvavr{155}{187}{48}}
        \\
        \hdashline\noalign{\vskip 0.1cm}
        
        \href{https://www.theverge.com/}{The Verge} & 2011-2022 
        & 720 \newline {\footnotesize \smvavr{89}{473}{158}} 
        & 236 \hfill {(33\%)} \newline {\footnotesize \smvavr{61}{160}{15}}
        & 175 \hfill {(24\%)} \newline {\footnotesize \smvavr{53}{114}{8}}  
        \\
        \hdashline\noalign{\vskip 0.1cm}
        
        \href{https://www.wired.com/}{WIRED} & 2010-2022 
        & 34000 \newline {\footnotesize \smvavr{5345}{17319}{11516}} 
        & 22940 \hfill {(67\%)} \newline {\footnotesize \smvavr{3954}{11560}{7426}}   
        & 1489 \hfill {(4\%)} \newline {\footnotesize \smvavr{921}{409}{159}}  
        \\ \midrule
        
        \textbf{Total} & 1997-2022 
        &  42405 \newline {\footnotesize \smvavr{7968}{20148}{14289}} 
        &  26628 \hfill {(63\%)} \newline {\footnotesize \smvavr{5503}{12855}{8270}} 
        &  2616 \hfill {(6\%)} \newline {\footnotesize \smvavr{1498}{840}{278}}  
        \\
        \bottomrule
        \end{tabular}
    \end{center}
\end{table*}
 }{  \definecolor{smvavrcolor}{HTML}{9e9e9e}
\newcommand{\smvavr}[3]{\textcolor{smvavrcolor}{#1$\cdot$#2$\cdot$#3}}

\begin{table}
    \small
    \caption{Online sources for retrieving articles on the three technology domains in our technical evaluation: Social Media (SM), Voice Assistants (VA), and Virtual Reality (VR). The table shows the percentage and total number of relevant articles that contain consequences after each filtering step.}

    \label{tab:title-classifier}
    \vspace{-0.2cm}
    \begin{tabular}{>{\raggedright\arraybackslash}p{1.7cm} >{\raggedright\arraybackslash}p{2cm} >{\raggedright\arraybackslash}p{1.8cm} >{\raggedright\arraybackslash}p{1.5cm}}
        \toprule
        \textbf{News Source} & \textbf{Retrieved Articles \newline {\footnotesize \smvavr{SM}{VA}{VR}}} & \textbf{Title Filter \newline {\footnotesize \smvavr{SM}{VA}{VR}}} & \textbf{Content Filter \newline {\footnotesize \smvavr{SM}{VA}{VR}}} \\
        \midrule
        \href{https://www.technologyreview.com/}{MIT Tech Review \newline \footnotesize{1997-2022}} & 3433 \newline {\footnotesize \smvavr{1686}{957}{790}}  & 1957 (57\%) \newline {\footnotesize \smvavr{1082}{563}{312}}   & 519 (15\%) \newline {\footnotesize \smvavr{349}{116}{54}} \\
        \hdashline[1.5pt/5pt]\noalign{\vskip 0.1cm}
        \href{https://techcrunch.com/}{TechCrunch \newline \footnotesize{2005-2022}} & 3975 \newline {\footnotesize \smvavr{748}{1502}{1725}} & 1330 (33\%) \newline {\footnotesize  \smvavr{337}{538}{455}}  & 390 (10\%) \newline {\footnotesize \smvavr{155}{187}{48}} \\
        \hdashline[1.5pt/5pt]\noalign{\vskip 0.1cm}
        \href{https://www.theverge.com/}{The Verge \newline \footnotesize{2011-2022}} & 720 \newline {\footnotesize \smvavr{89}{473}{158}}  & 236 (33\%) \newline {\footnotesize \smvavr{61}{160}{15}} & 175 (24\%) \newline {\footnotesize \smvavr{53}{114}{8}} \\
        \hdashline[1.5pt/5pt]\noalign{\vskip 0.1cm}
        \href{https://www.wired.com/}{WIRED \newline \footnotesize{2010-2022}} & 34000 \newline {\footnotesize \smvavr{5345}{17319}{11516}}  & 22940 (67\%) \newline {\footnotesize \smvavr{3954}{11560}{7426}}  & 1489 (4\%) \newline {\footnotesize \smvavr{921}{409}{159}} \\

        \midrule
        \textbf{Total} \newline {\footnotesize 1997-2022} & 42405 \newline {\footnotesize \smvavr{7968}{20148}{14289}} & 26628 (63\%) \newline {\footnotesize \smvavr{5503}{12855}{8270}} & 2616 (6\%) \newline {\footnotesize \smvavr{1498}{840}{278}} \\
        \bottomrule
    \end{tabular}
    \vspace{-0.3cm}
\end{table}
 } 

%% file: sections/04-study.tex
\section{Study 1}
\label{user_evaluation_1}
Our first study evaluated \name's overall usefulness for researchers who are experts in a technology domain that is currently covered in \name. Specifically, our study investigated whether a catalog of \consequences in \name could enhance their awareness of potential impacts within their general CS subarea. The study design was guided by three research questions:

\begin{enumerate}
    \item[\textbf{RQ1}] Does \name support CS researchers in discovering undesirable consequences beyond their prior knowledge and beyond searching on the internet?
    \item[\textbf{RQ2}] How do researchers perceive \name's usefulness for discovering \consequences in their area?
    \item[\textbf{RQ3}] How and when do researchers imagine using \name during the research and development process?
\end{enumerate}

\subsection{Methods}
We chose a within-subjects design to answer whether \name helps researchers gain insights into undesirable consequences in their technology domain, beyond what they already know and beyond what they may discover through an online search. 

\subsubsection{Participants}
We recruited nine computer science (CS) researchers (7 male, 2 female) aged 24-41 ($\mu = 27.55$, $\sigma = 5.66$) years old through personal connections and institutional Slack channels. Our participants are based in the US and fluent in English. To ensure participants actively worked on cutting-edge technologies, our selection criteria required participants to be CS researchers in academia or industry and develop technologies in the areas of social media (SM), voice assistants (VA), or virtual reality (VR) (3 from each) and to have authored at least one peer-reviewed publication in one of these technology domains. Importantly, our participants were not new to the topic of \consequences. Two participants work on ethics-related topics (AI fairness and identifying security and privacy issues in VR). Of the 9 participants, 7 are Ph.D. students, 1 is a postdoctoral researcher, and 1 is a research scientist at a non-profit research institute. Seven participants previously worked for large multinational technology corporations relevant to their research fields. We met with 5 participants over Zoom (P3, P5-6, P8-9) and 4 in person (P1, P2, P4, P7).

\subsubsection{Procedure}
Each study session started by obtaining consent, followed by a brief introduction of \emph{undesirable consequences of technology}. All participants used their own devices for the study. The recruitment email did not detail the study procedure to prevent them from preparing in advance and confounding our baseline conditions.
Each participant took part in three ordered conditions:
\begin{enumerate}[wide, labelwidth=!, labelindent=0pt]
    \item \aptLtoX[graphic=no,type=html]{ {\colorbox{gray!30}{\textsc{Know}}}\xspace }{ \cone } (\textbf{Prior Knowledge Baseline}): Participants were asked to think about and list any \consequence of digital technologies in their domain of expertise (one of SM, VA, or VR), solely based on their prior knowledge.
    \item \aptLtoX[graphic=no,type=html]{ {\colorbox{ctwo!20}{\textsc{Search}}\xspace} }{ \ctwo }  (\textbf{Online Search Baseline}): Participants were asked to search any online resources of their preference (e.g., Google Search, Google Scholar, Semantic Scholar) to add to the previously generated list of \consequences.
    \item \aptLtoX[graphic=no,type=html]{ {\colorbox{pink!40}{\name}\xspace} }{ \cthree }  (\textbf{\name System}): Participants were asked to interact with \name and list any additional consequences that they did not mention in either of the prior conditions. Participants were provided the URL for \name without any additional instructions on how to use it.
\end{enumerate}

We chose the two baseline conditions because these are plausible alternatives for exploring \consequences, which technologists desire to do but not yet commonly practice~\cite{Do2023ThatsIB}. 
Each condition was limited to 15 minutes to allow for comparable outcomes across conditions and to keep the study duration to at most one hour. None of the participants reached this limit in any of the conditions. Participants were alerted about the remaining time after 10 minutes.  They were also informed that they could jump to the next condition when they could not think of, or find, more \consequences. While participants were encouraged to think of as many unique \consequences as possible, all of them eventually ran out of ideas and switched to the next condition. 
   
Each study session ended with a semi-structured interview eliciting feedback on \name. The interview asked questions about the perceived usefulness and challenges of \name as well as how participants would integrate \name in their research process and what future use cases they could envision for \name. 

\subsubsection{Analysis}
To answer whether \name supports \developers in naming \consequences beyond their prior knowledge and an online search (\textbf{RQ1}), we analyzed the number of unique \consequences listed during each condition. Consequences were considered as unique if they considerably differed in detail---for example, in terms of different ``aspects'' like privacy issues due to recording vs. data leaks or specific examples of \consequences affecting certain populations in different ways. We also required consequences to be either already existing or reasonable (e.g., they could be anticipated but had to be realistic). 
We counted the results by the conceptual distinction made by the participant barring repetition or similar incidents (e.g., virtual reality might cause nausea or motion sickness are counted as one unique idea).
Two authors independently coded and counted participants' unique consequences. During analysis, the three conditions were randomized to prevent confirmation bias towards our system, \name. To assess the category consistency, we calculated the inter-rater reliability of the categories using Cohen's Kappa $\kappa = 85.33\%$. For the 21 responses out of 155 that the authors disagreed on, two authors discussed and decided on the final aspect. The anonymized user data can be found in the Supplementary Materials.

To answer \textbf{RQ2} and \textbf{RQ3}, we conducted a thematic analysis of the semi-structured interviews. First, two authors individually reviewed, and conducted open coding for, three interviews. Next, three authors met to consolidate and create the first draft of the codebook. Finally, two authors independently coded the remaining interviews and refined the codebook, which was discussed with the full research team. In line with prior suggestions~\cite{mcdonald2019reliability}, we did not calculate an inter-rater reliability score for the interview codes because our goal was to uncover more themes.

\paragraph{Researchers' Positionality}
Our motivations and perspectives for designing, developing, and evaluating \name are shaped by our academic and professional roles as US-based CS researchers at an R1 university. With backgrounds in HCI, NLP, Software Engineering, and Computing Ethics, we had many discussions about supporting and incentivizing researchers in various subfields of computer science to learn about and anticipate \consequences of technology that all authors have experienced in the past. All of the authors have had prior research experience (as faculty, interns, or research scientists) at other universities and in industry labs, which has influenced our thinking of how the inertia to consider \consequences could be overcome. While we are ultimately technologists, we see \name not as a complete solution to the problem, but as being in a supportive role that needs to be combined with incentives and structural changes in academia. 

%% file: sections/05-result.tex
\begin{figure}
\centering
\hfill
  \begin{minipage}[t]{0.46\columnwidth}
        \centering
        \includegraphics[width=\textwidth]{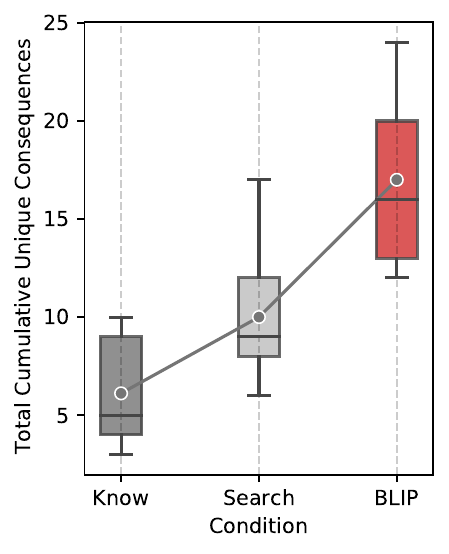}
        \vspace{-0.5cm}
        \caption{Cumulative unique number of consequences in each condition. The line represents means of unique consequences between the three conditions.}
        \Description{This box and whisker plot shows the cumulative unique number of consequences produced for each condition in the user study. On the x-axis are the three conditions: Know, Search, and BLIP. The y-axis is the cumulative unique consequences. We can observe an increase in consequences from left to right (Know to BLIP conditions).}
        \label{fig:consequence-count}
    \end{minipage}
    \hfill
    \hspace{0.01\columnwidth}
    \begin{minipage}[t]{0.46\columnwidth}
        \centering
        \includegraphics[width=\textwidth]{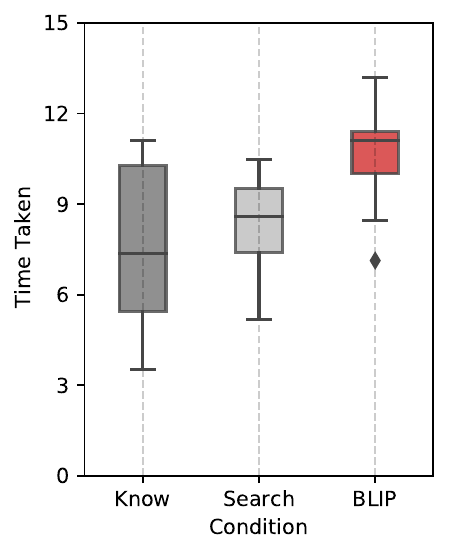}
        \vspace{-0.5cm}
        \caption{Total time taken for each of the three conditions separately (in minutes). }
        \Description{This box and whisker plot shows the time taken for each condition in the user study. On the x-axis are the three conditions: Know, Search, and BLIP. The y-axis is the time taken for each condition. We can observe that BLIP has the highest time taken.}
        \label{fig:consequence-time}
    \end{minipage}
\hfill
\vskip 0.3cm
\hfill
 \includegraphics[width=\columnwidth]{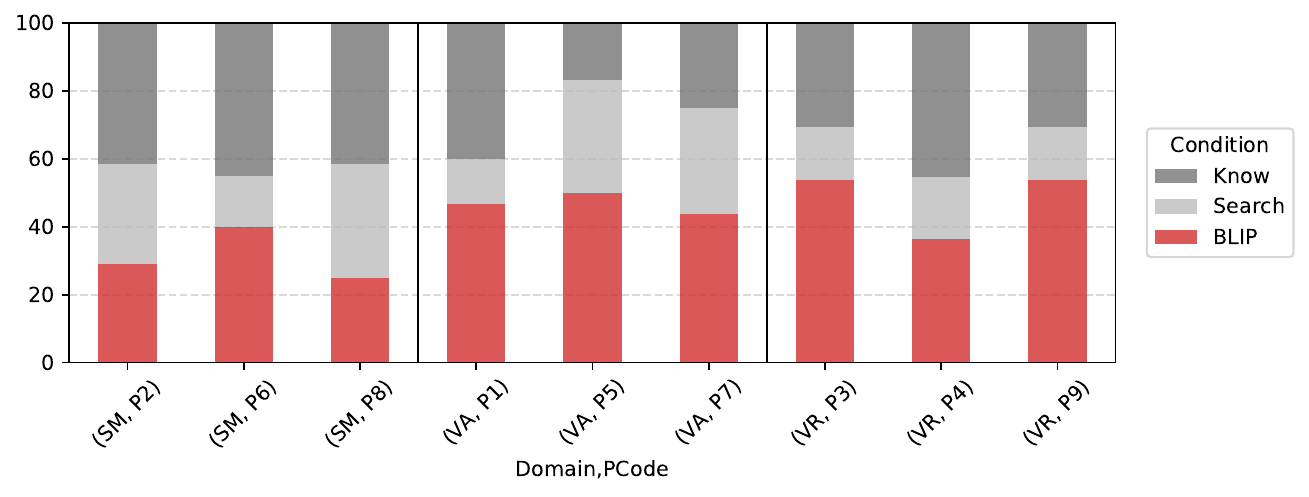}
    \vspace{-0.5cm}
     \caption{Percentages of total consequences by each participant for each condition of our study by technology domain.}
     \Description{This bar plot shows the percentages of total consequences by each participant for each condition of our study. There are nine bars, each of which have the same height. The proportion of three conditions are indicated in each bar. We can observe a higher proportion of consequences in the Know condition for the social media condition, roughly speaking.}
    \label{fig:results}
\end{figure}

\subsection{Results}
\label{user_evalution_1_result}
We report our findings in the following two sections. 
\vspace{-0.2cm}
\subsubsection{\textbf{\name allows participants to find additional and diverse \consequences (RQ1)}}
Our quantitative analysis showed that \name supported participants in discovering \consequences beyond their prior knowledge and beyond what they found during an Internet search. In the  \aptLtoX[graphic=no,type=html]{ {\colorbox{gray!30}{\textsc{Know}}}\xspace }{\cone} condition, participants listed an average of 6.11 (SD=2.80) unique \consequences (\autoref{fig:consequence-count}). For example, participants mentioned potential implications ranging from an ``echo chamber'' in social media to sensitive bio-metric information used for advertisement in virtual reality. They mentioned reading about these examples in the news or research papers, in addition to sometimes citing their own work. Participants spent an average of 7.52 minutes (SD=2.68) on this first condition as shown in \autoref{fig:consequence-time}. All participants switched to the next condition before the 15-minute limit was reached. 

The \aptLtoX[graphic=no,type=html]{ {\colorbox{ctwo!20}{\textsc{Search}}\xspace} }{\ctwo} condition only resulted in an average of 3.88 additional \consequences (SD=1.83). Overall, participants spent more time in this condition than in the \aptLtoX[graphic=no,type=html]{ {\colorbox{gray!30}{\textsc{Know}}}\xspace }{\cone} condition (\aptLtoX[graphic=no,type=html]{ {\colorbox{gray!30}{\textsc{Know}}}\xspace }{\cone}: $M=7.52$ minutes, $SD=2.68$; \aptLtoX[graphic=no,type=html]{ {\colorbox{ctwo!20}{\textsc{Search}}\xspace} }{\ctwo}: $M=8.23$ minutes, $SD=1.96$)
The difference is mainly due to the time required for searching for, and reading through, information online. The fact that they found fewer ideas on average in \aptLtoX[graphic=no,type=html]{ {\colorbox{ctwo!20}{\textsc{Search}}\xspace} }{\ctwo} makes intuitive sense because participants searched for consequences similar to or building upon what they already thought of in the \aptLtoX[graphic=no,type=html]{ {\colorbox{gray!30}{\textsc{Know}}}\xspace }{\cone} condition. Our observations suggest that searching for resources online, such as through Google Search, Google Scholar, or Semantic Scholar, is not well-suited for finding \consequence, owing to the fact that it often necessitates prior knowledge on what to search for. While participants used different combinations of search engines (5 participants only used Google Search, whereas 4 others used both Google Search and Google Scholar or Semantic Scholar), they were unable to find many new consequences. As P8 noted, \emph{``all of the titles on Google said the same things ..., so I had to open them to see [the content], which is tiring."} P3 even had trouble finding the right keyword to search for the content, stating that \emph{``I don't know if there just isn't so much work about [safety issues of VR] or I just searched with the wrong keywords''}. All participants switched to \name before the 15 minutes ended, suggesting that they had exhausted ways for searching for information about \consequences. 
 
With \aptLtoX[graphic=no,type=html]{ {\colorbox{pink!40}{\name}\xspace} }{\cthree}, participants were able to add an average of 7.00 \consequences ($SD=1.65$) not listed before. P3 summarized their experience by saying \emph{``a lot of ideas just came to me that I otherwise would never have thought of.''} As an expert on social media, P2 mentioned an additional 7 ideas when using \name compared with 17 ideas in the prior baseline conditions (\aptLtoX[graphic=no,type=html]{ {\colorbox{gray!30}{\textsc{Know}}}\xspace }{\cone}: 10, \aptLtoX[graphic=no,type=html]{ {\colorbox{ctwo!20}{\textsc{Search}}\xspace} }{\ctwo}: 7). For instance, they commented that as social media platforms grew, governments can easily censor the population by deleting controversial topics, a consequence which they had ``\textit{totally missed}'' in the previous conditions.  

While participants were able to expand their list of \consequences using \name, we found that participants spent, on average, more time with \aptLtoX[graphic=no,type=html]{ {\colorbox{pink!40}{\name}\xspace} }{\cthree} ($M=10.52\text{ minutes}, SD=1.85$) than in the other two conditions (see \autoref{fig:consequence-time}). Our observations and analysis of post-study interviews suggest that \name was perceived as engaging and as a tool that continuously led to new insights. For example, participants commented that \name enables them to find examples of \consequences quickly (and especially more quickly than in the \aptLtoX[graphic=no,type=html]{ {\colorbox{ctwo!20}{\textsc{Search}}\xspace} }{\ctwo} condition). P1 mentioned that they were completely oblivious to the time they spent exploring \name until we reminded them after 10 minutes in the \aptLtoX[graphic=no,type=html]{ {\colorbox{pink!40}{\name}\xspace} }{\cthree} condition. P9 described the interface as \emph{``addictive''} and that they would like to keep re-visiting \name in the future. This suggests that, unlike in the \aptLtoX[graphic=no,type=html]{ {\colorbox{ctwo!20}{\textsc{Search}}\xspace} }{\ctwo} condition, participants felt like they wanted to spend more time exploring \consequences.  
 
A detailed breakdown of the percentage of unique consequences listed in each condition by each participant can be found in \autoref{fig:results}. Based on the percentage of new consequences added in each condition, \name appears to be most useful for VR, VA, and SM in that order. Such an order could be the result of the different eras in which these technologies were introduced---the \consequences of VR and VA are just becoming apparent, whereas those of SM have been known for some time and are regularly discussed in the news. Participants working on social media were therefore able to cover many consequences in the \aptLtoX[graphic=no,type=html]{ {\colorbox{gray!30}{\textsc{Know}}}\xspace }{\cone} condition. 

\begin{figure}
    \centering
    \includegraphics[width=\columnwidth]{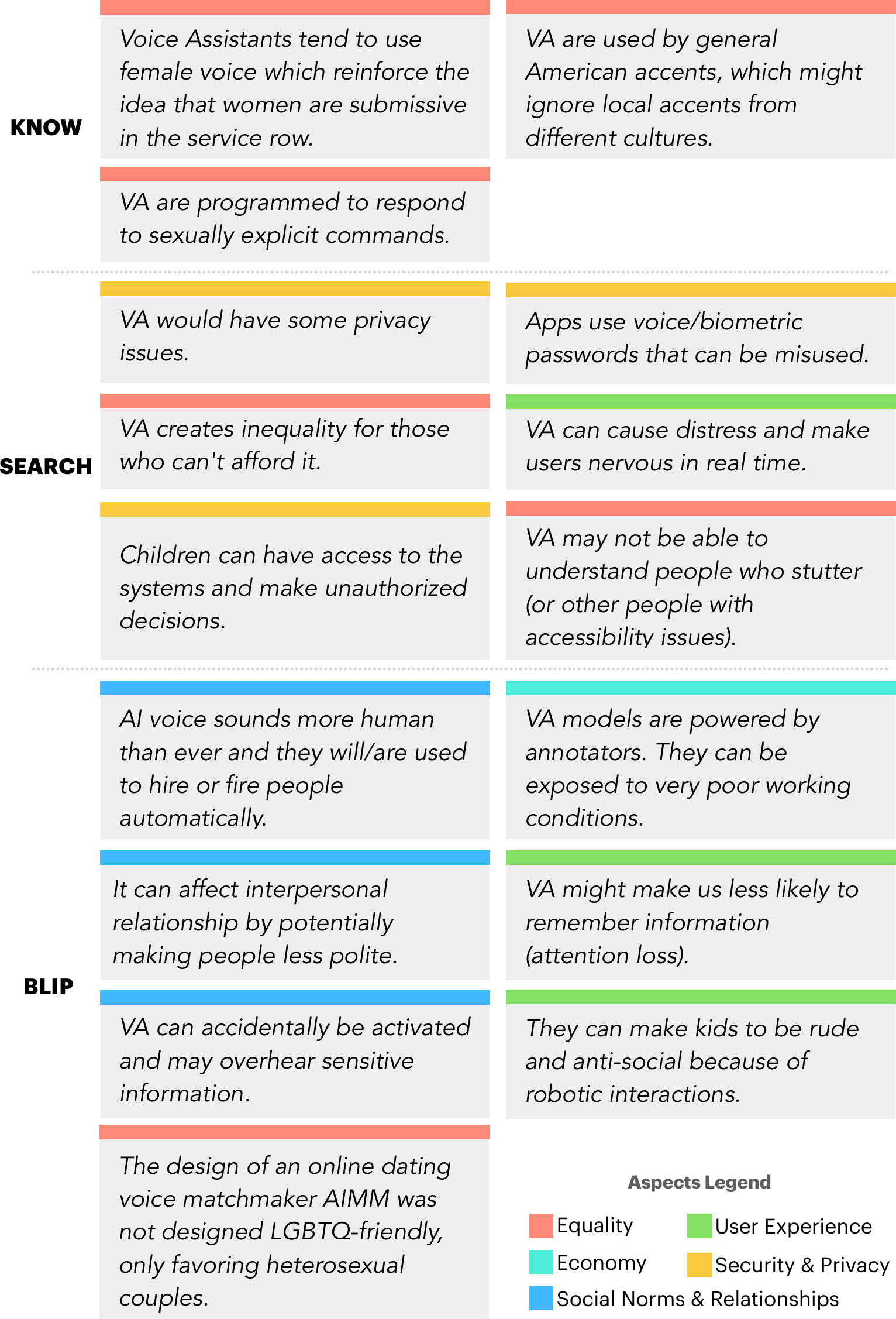}
    \vspace{-0.3cm}
    \caption{Quotes from P5's reporting of \consequences of voice assistants and the associated aspects of life in the three conditions of our user study.}
    \Description{Quotes from P5's reporting of undesirable consequences of voice assistances and the associated aspects of life in the three conditions in our user study. We can observe the three consequences in the Know condition with the same aspect of life ("equality" in red), 6 consequence in 3 categories in the Search condition ("security & privacy", "equality", and "user experience") and 7 categories in 4 categories in the BLIP condition ("Social Norms & Relationships", "User Experience", "Equality", "Economy".)}
    \label{fig:study_example}
    \vspace{-0.6cm}
\end{figure}

While the number of \consequences participants were able to think of using \name is encouraging, we also analyzed whether \name helped participants think of more diverse instances than in the \aptLtoX[graphic=no,type=html]{ {\colorbox{gray!30}{\textsc{Know}}}\xspace }{\cone} and \aptLtoX[graphic=no,type=html]{ {\colorbox{ctwo!20}{\textsc{Search}}\xspace} }{\ctwo} conditions. For this analysis, we manually categorized each of their responses into one of the 10 aspects of life in Section \ref{sec:categorization}. The consequences can be categorized into an average of 4.11 aspects in the \aptLtoX[graphic=no,type=html]{ {\colorbox{gray!30}{\textsc{Know}}}\xspace }{\cone} condition, 2.78 in the \aptLtoX[graphic=no,type=html]{ {\colorbox{ctwo!20}{\textsc{Search}}\xspace} }{\ctwo} condition, and 5.00 in the \aptLtoX[graphic=no,type=html]{ {\colorbox{pink!40}{\name}\xspace} }{\cthree} condition. To exemplify how participants broadened their list of \consequences across the three conditions, ~\autoref{fig:study_example} shows that P5 first fixated on \consequences related to Equality in the \aptLtoX[graphic=no,type=html]{ {\colorbox{gray!30}{\textsc{Know}}}\xspace }{\cone} condition. Searching for consequences broadened their list to three different aspects (Security \& Privacy, Equality, and User Experience). It was only when they started using \name that they additionally thought of impacts on Economy and Social Norms \& Relationships, in addition to others. As we will show in our qualitative results in the next section, several participants echoed our finding that \name helps users diversify their list of \consequences. 


Altogether, our findings indicate that \name supports users in discovering \consequences beyond their prior knowledge and searching online, affirming our first research question.

\subsubsection{\textbf{How participants perceived \name's usefulness and how it would be used (RQ2 and RQ3)}}
To answer our second and third research questions, we present three high-level themes as revealed by our qualitative analysis. 

 \textbf{\name is useful for learning about \consequences and reflecting on their own experiences.}
Our analysis revealed that participants found \name improves their ability to \emph{``think outside the box''} and made them aware of \consequences that they \emph{``had never considered''} [P3]. Many participants suggested that a tool like \name is essential and wished it had existed earlier. For instance, P7, a Ph.D. student who studied conversational AI, said they wished they had looked through the examples or \emph{``at least thought about these societal issues''} in their first project. They regretted working on an issue with a \emph{``very poor understanding of the social implications.''} They explained that ignoring societal ramifications can be a bigger problem for technical fields disconnected from end-users such as NLP because \emph{``people have established benchmarks and evaluation metrics, [so] you [researchers] can work on the task without having any idea of how it's used in real life.''} P1 echoed that thinking of \consequence is \emph{``difficult''} if researchers do not work specifically on fairness or accountability issues. Because \name pre-processes the real-world examples, P7 found it to be \emph{``easier and faster''} to learn about \consequences compared to relying on their prior knowledge or online search.

Participants also stated that \name's aspect categories are useful for broadening their knowledge of \consequences. 
 For example, P6, who authored over 5 papers on mental health issues on social media, extensively discussed \consequences such as how social media can cause people to feel lonely, show \emph{``scary''} images that can affect users' mental stage, and idealize beautified photos that make teenage girls feel bad about themselves. When searching online, they continued focusing on health-related risks such as how social media can increase depression and schizophrenia. \name expanded their discussion to other issues on privacy, economy, online bots, limitation of freedom of speech, and interpersonal relationships. In the end, P6 commented on the benefit of having different aspect categories: \emph{``I do feel like it's very nice to get exposed to a wide variety of topics... I feel like this would be great to anticipate [\consequences of] a new social media product.''}
 
 We observed a similar pattern for P1, P3, P5, P8, and P9, all of whom focused on aspects that are related to their research areas in the baseline conditions and only included more diverse issues once they started using \name. 
 Note that we explicitly asked participants to list diverse \consequences until they exhausted their ideas, at which point they could switch to the next condition. The fact that only three participants generated consequences across several aspects of life from the beginning suggests that many researchers and developers may fixate on specific issues. The diversity of \consequences in \name was useful for gaining insights into additional aspects and overcoming this fixation. 
  
Participants also noted that the examples in \name inspired them to think of their own prior experiences or additional \consequences. For instance, for P1, a summary on one card (\emph{``People have become more reliant on machines to do tasks that they are capable of doing themselves.''}) inspired them to recall their own experience with voice assistants: \emph{``[I] deliberately speak in a way that it will be easier for [the] machine to understand.''} They recalled that the interaction \emph{"fundamentally changed my behaviors."} Every participant mentioned at least one incident (\emph{``This made me think of ...''}). As P8 said after using \name, ``\emph{I felt that I was exposed to a bunch of possible directions that you can see. I feel that I'm learning a lot.}'' 
  
\textbf{\name can be useful for brainstorming \consequences, writing ethics statements, and for different stakeholders.}
We found that participants appreciated \name for helping them brainstorm \consequences of their own innovations, including when writing an ethics statement or introduction for a paper. P4, who studied security issues on VR devices, suggested that \name could be useful for finding arguments and citations for their publication discussing the \consequence of VR. P1-2, P4, and P7-8 all mentioned that they perceive \name to be useful to get ideas for ethics statements or the introduction of their papers.  P4 was hopeful that \name could play a role in establishing a \emph{``brainstorming phrase''} for CS researchers to determine the research questions to address. Similarly, P7 indicated that they would use a system like \name to think about \consequences of a new topic early on, stating that \emph{``it will be very useful for brainstorming and will [help me] process a lot of information faster.''} 
  
Participants also felt that \name could be useful for a variety of stakeholders, not just for themselves. Several participants suggested that it would provide a good introduction to \consequences for the general public, researchers, or developers who are new to a particular field. As P9 mentioned: \emph{``I think in order to get a sense of the full paradigm of VR, you have to have a tool like this because you can't just read a bunch of disconnected articles about this.''} P1 made a similar comment that users could quickly browse through a \emph{``broad spectrum of issues''}. P3 also appreciated that \name can enable readers to \emph{``quickly skim through the summaries''} without delving into each online resource.

In summary, the results of our first study suggest that \name supports the discovery of more and more diverse \consequences relevant to specific technology domains compared to prior knowledge and an online search (\textbf{RQ1}). \name was perceived useful to learn about about \consequences and reflect on their own experience (\textbf{RQ2}). In addition, participants found that \name can be useful for brainstorming \consequences, writing ethics statements, and for different stakeholders (\textbf{RQ3}).

\vspace{-0.3cm}
\subsubsection{Participant Feedback and System Changes}

Our interviews revealed several opportunities for improving \name, which we subsequently implemented. Specifically, participants (P2, P4-5) suggested that in addition to online articles, it would be helpful to also have academic articles available---in particular those that uncover and discuss \consequences. We therefore added the functionality to parse and summarize academic papers, adding the last twenty years of papers from the CHI proceedings (2003-2023) as a data source. In \name's GUI, users can filter whether they want to see all data sources, only academic papers, or only articles from online magazines.  
We also made minor changes to the GUI based on their feedback, such as changing the appearance of the summaries and buttons. 
Additionally, we included the Llama2 open-source model in the backend as more large language models become available~\cite{touvron2023llama}.

%% file: sections/05.5-new_study.tex
\section{Study 2}

Our first study showed that \name can indeed increase researchers' awareness of \consequences in their CS subfield compared to relying on their prior knowledge or searching online. What remained unanswered was whether \name is  useful for collecting, considering, and acting on  \consequences relevant to specific projects users work on. Our second study, therefore, focuses on the following research question (\textbf{RQ4}): To what extent is \name useful and actionable for \rr{users'} \emph{own projects}? We study this question both objectively (i.e., whether they can find \consequences that are relevant to their projects) and subjectively (whether they perceive \name as useful and actionable). 

\subsection{Methods}

\subsubsection{Participants}
We recruited six CS researchers (4 male, 2 female) aged 23-26 ($\mu=25.00$, $\sigma=2.00$) years old through personal connections and institutional Slack channels. Our participants are based in the US and fluent in English and all are currently Ph.D. students with experience in the technology industry through internships or prior work experience. The six researchers work in Computer Vision, Vision Language Models, Mobile Technology, Computational Biology, Robotics, and Ubiquitous Computing. To protect their anonymity, we only refer to their general research directions and avoid discussing the specifics of any project. 

\subsubsection{Procedure}
We met participants over Zoom (P1-2, 4-5) and in person (P3, 6).
Before each session, we used \name's bulk-import functionality (see Section \ref{extension}) to filter, summarize, and categorize new content specific to the six CS subfields that participants' work contributes to. 
\rr{We added all content on \name, including a list of domain keywords and the imported articles, before the user studies to ensure that participants could explore consequences in their own subfields (e.g., Ubiquitous Computing).}
At the beginning of the study, we asked participants to describe their current project. 
They were then explicitly instructed to use \name to bookmark as many \consequence as they may find relevant to their own projects as if they were trying to establish a comprehensive list.
Participants could open the articles in \name if they were interested. 
To approximate a real-world usage scenario and avoid making participants feel observed, the experimenter left the study session until participants messaged the authors that they had gained a sufficient overview of the relevant \consequence. Participants were told that they had a maximum of 30 minutes to explore \name. 

At the end of the session, we conducted a brief interview with each participant. Specifically, we asked participants to share their screens and explain they found the bookmarked articles relevant to their \rr{projects}. This was done to understand how participants would use the \consequences described in the bookmarked articles and whether they were actionable. We also asked participants which features in \name they liked and what may encourage or prevent them from using \name, as well as what improvements they would recommend. 
After completing the session, participants were sent a link to a post-study survey asking them about the usefulness of \name for considering \consequences in their own projects. All responses were anonymous to reduce potential response bias. (See \autoref{fig:survey_results} for specific questions.)

\subsubsection{Measures}
We collected the total time participants used \name, the number and content of bookmarked articles, and responses to the survey questions. We recorded the post-study interviews for qualitative analysis. Similar to Study 1, two authors individually reviewed and conducted open coding on the initial two interviews. The two authors then independently coded several more interviews and refined the codebook. We did not calculate the inter-rater reliability for the codes to discover more emergent themes~\cite{mcdonald2019reliability}.

\subsection{Results}

Participants spent an average of 13.11 minutes ($SD=4.72$) browsing through \name and bookmarked an average of 7.67 articles ($SD=3.94$, $\text{Min}=5$, $\text{Max}=13$) relevant to their research articles (See Supplementary Materials for examples of bookmarked articles). The result suggests that participants can find \consequences through \name that are relevant to their own projects and that they can do so in less than 15 minutes. 

\begin{figure*}
    \centering
    \includegraphics[width=0.95\linewidth]{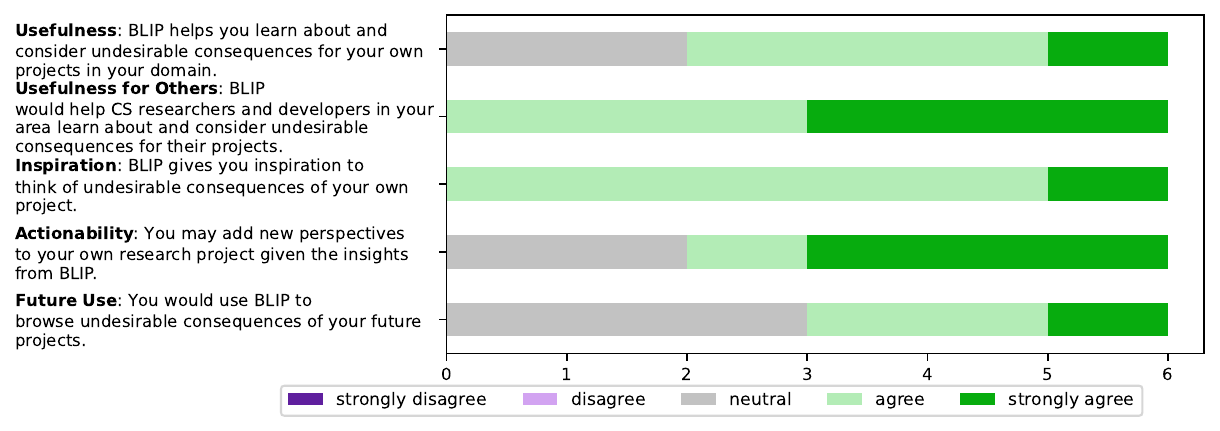}
    \vspace{-0.4cm}
    \caption{Post-study Survey Responses in Study 2. The x-axis shows the number of participants (N=6)}
    \Description{This horizontal bar chart shows the number of participants who strongly disagree, disagree, neutral, agree, and strongly agree on five questions (i.e., Usefulness, Usefulness for Others, Inspiration, Actionability, and Future Use.)}
    \label{fig:survey_results}
\end{figure*}

Our post-study survey results showed that participants perceived \name generally as useful and actionable 
(see Fig.~\ref{fig:survey_results}).
Specifically, four of our six participants agreed or strongly agreed that \name is useful for learning and considering \consequences relevant to their own project (two were neutral). 
All participants agreed or strongly agreed that \name would be useful for others working in their area. 
\name was also seen as an inspiration to think of \consequences of participants' projects (five agreed, one strongly agreed). 
Finally, three participants agreed or strongly agreed that \name provides them with new perspectives for their project, suggesting its actionability (two were neutral) and three participants would use it for future projects (three were neutral). 

\subsubsection{Interview Results}
The interviews with participants helped to shed further light on these results. 
First, \textbf{participants found and bookmarked summaries that were relevant to their projects}. For example, P6 who worked on ubiquitous computing, bookmarked a mixture of online articles and CHI papers. Their current project was focused on building smart electrocardiogram (ECG) systems, so they bookmarked summaries that were particularly relevant to ECG and wearables. One summary suggested that wrist devices lack accuracy when people are mobile or if the tasks require physical activity; another one described ECG biometrics as lacking robustness when deployed in the wild. 
During the interview, when asked about the relevance of the bookmarked summaries for their project, they pointed out an article titled \emph{What to Put on the User: Sensing Technologies for Studies and Physiology Aware Systems}. P6 said that \emph{``the source seems to cover challenges and the different physiological signals that we can use for wearable technologies [...] I think it will be helpful to know [the signals] besides what I'm using right now and be aware of the issues in the future.''}  P6 suggested \name made them look at potential real-world implications, instead of merely finding, replicating, and improving \emph{``lower-level and technical systems from the very beginning.''} They commented that \name offers a very different perspective to think of their research in ubiquitous computing. They later notified the authors that they reviewed the bookmarked list after our study and read the articles in detail.

Similarly, P2, who works on vision language models, bookmarked an article describing that voice assistants may have the potential to reduce creativity due to overpersonalization. When asked about this article in the interview, P2 mentioned that their research project includes a \emph{``component to add user's preference and context input to improve the model''} but they did not think that this would have any consequence at all. P2 continued that \emph{``overpersonalization can be a problem, right? Maybe we [our team] should think more about this feature.''} 
P4, who worked on a computational biology project and \emph{``did not think about ethics too much,''} commented that they often focus on the technical aspect but had never read so many perspectives about potential \emph{``risks''} in the real world before using \name. They also pointed to articles that they thought relevant to their projects about the \consequence of gene therapy after searching keywords such as ``gene therapy'' and ``t-cell receptors''.

Second, the interviews showed that \textbf{participants generally perceived \name as useful and actionable}. For example, P5 mentioned that they often read magazines such as The Verge, and \emph{``[using] \name feels like reading a more relevant news page.''} P5 mentioned that \name helps them to rethink ways to alleviate the potential consequences of robots. They found an article about the impact of robots on economic inequality and commented that they had not thought much about this issue since automation seems always \emph{``the way to go.''} They had also bookmarked two articles about potential solutions titled \emph{``One way to get self-driving cars on the road faster: let insurers control them''} and \emph{``EU proposals would classify robots as electronic persons.''} Inspired by the two articles, P5 proposed how one could address the problem, suggesting they could implement \emph{``subscription service where robots just act as an agent so that everyone can control the robots and receive the payment.''} P5 acknowledged that this \rr{\emph{``of course, is a very naive solution,''}} but \emph{``I should start think more about these issues''} as \name have exposed many articles to them. 

%
Third, the interviews explained why \textbf{some, but not all,  participants thought they would integrate \name into their work in the future}. Several participants stressed the importance of having \name. For example, P2 stated \emph{``I think using this tool is particularly useful for ethical consideration, like [the] ethical statements section in the ACL papers because most of [the] time researchers don't know what concrete examples to write for those sections, and this provides good resources.''} P2 continued to say that \emph{``a lot of the resources are from trendy topics on magazines and the venues that I'm not familiar with, [un]like ACL or other pure machine learning conferences.''} P2 later told the authors that they would try to use \name for their next paper submission. P1 also mentioned that they would use \name to find relevant literature, which they would like to include in the introduction or related work sections. P3 also stressed having such a catalog of consequences is \emph{``necessary''} to keep up with the related work and how the media portrays some issues. According to P3, this is in contrast to Google Scholar feed which only recommends technical work for them.

Other participants who did not think they would use \name in the future (P4-6) viewed the task of addressing undesirable consequences as tangential to their primary focus on technical development. Their comments suggest that a tool alone will not change people's perceived responsibility for actively considering \consequences. 
For example, P6 mentioned that they learned \emph{``a lot [about] social impact''} by looking at \name, but their research \emph{``primarily focuses on building a technical system.''} They were not sure whether they should be the person thinking about \consequences. 
P5 mentioned that most of their research considered \emph{``achieving human-level dexterity of robotics,''} but most of the \consequences in \name are very \emph{``socio-economical.''} They did not think that they are at the right position or qualified to think about and address these issues.

Finally, \textbf{participants suggested opportunities for making \name more useful for them}. 
As in the previous study, several participants suggested that expanding \name's catalog to other sources may provide additional inspiration. 
For example, P3 mentioned they found \emph{``some cool articles''} and relevant CHI papers, but they were interested in \emph{``how many papers in my field [Mobile Technology] and outside CHI revenue portrayed these problems.''} They suggested including a list of publication venues and workshops across disciplines but realized that it probably went beyond the \name's scope.

P5 commented on improving \name's actionability by suggesting that the system would greatly benefit researchers if it could include a \emph{``solution''} to the \consequences. 
In a similar vein, P1 suggested adding potential expert reviews or opinions. \emph{``I'm working on a very specific and new domain. I've talked about \consequence with my labmates but I'd be interested in hearing more experts to comment on their thoughts.''}

In summary, the interviews revealed that participants recognized the \name's potential and gained ideas of potential consequences that they would need to address in their projects. Participants were also envisioning \name 2.0, suggesting the need for tools that summarize the broad range of scientific literature from diverse fields on \consequences and aid in finding solutions. However, our results also showed that the onus of proactively addressing \consequence cannot rest solely on \name; integrating into existing research processes could further amplify its impact.

%% file: sections/06-discussion.tex
\vspace{-0.3cm}
\section{Discussion}

Our goal in this work was to evaluate whether providing 
CS researchers with an easily accessible catalog of undesirable consequences of digital technologies could improve their ability to learn about and consider adverse effects, as prior work had suggested~\cite{mcgregor2021preventing,Do2023ThatsIB, skirpanEthics2018}. To study this question, we developed \name, a web-based prototype that leverages language models to automatically derive \consequences from any given online article. \name addresses the difficulty of having a broad knowledge of potential \consequences, which, according to Merton, is ``the most obvious limitation to a correct anticipation.''~\cite{merton1936unanticipated} 

Our results show \name's potential for supporting CS \rr{researchers} in gaining awareness of a broad range of \consequences. 
In Study 1, \name supported researchers in finding more, and more diverse, \consequences of technology in their CS subdiscipline even after listing ones from prior knowledge and after searching online. When relying on their prior knowledge, we found that participants only thought of an average of 6.11 unique \consequences despite being experts in their technology domain. They were often stuck describing \consequences within one or two commonly known ``aspects'' (that were sometimes part of their research focus). This indicates that many researchers do not have a thorough and broad awareness of the \consequences within their technology domain. Intuitively, searching online might be a better option to explore additional \consequences to extend users' prior knowledge. However, our results showed that searching online only added an average of 3.88 \consequences and was perceived as a tedious approach. The fixation issue persisted when searching online, that is, participants mostly used search terms related to the \consequences they had already listed. 
This limitation underscores the insufficiency of traditional methods, as they often lead to a narrow focus, overlooking broader and potentially more impactful consequences.
Compared to the two baseline conditions, \name was able to support participants in listing and learning about \consequences that were often beyond the commonly known ones. \rr{To summarize these results, our study demonstrates that relying on prior knowledge and an online search, without any tooling support, is often perceived by the participants as tedious and insufficient for thinking about the \consequences of technology.}

 In Study 1, the qualitative responses further illustrate the most helpful parts of \name's design. We found that participants perceived \name's summaries of \consequences as beneficial for efficiently gaining an overview, while appreciating having access to the original articles to ensure information integrity. \name's categorization of different life aspects was seen as a motivating nudge for exploring consequences broadly---a finding that is in line with the results of our quantitative analysis. The result extends prior work in creativity and cognition that has found that providing a solution space using a set of dimensions breaks people's tendency to fixate~\cite{Ngoon2018InteractiveGT,siangliulue16:ideahound}: Providing a diverse set of \consequences can help technology experts consider societal implications broadly and reveal those that they would have otherwise not thought about. 

 Study 2 aimed to evaluate whether \name provides actionable information when freely using it to find potential \consequences relevant to researchers' specific projects, rather than to the whole field. We found that participants took less than 15 minutes, on average, to gather a set of consequences relevant to a specific research project and bookmarked an average of 7.67 unique \consequences during this time. While this second study was not designed to determine whether participants were able to \emph{comprehensively} find \consequences, participants' comments suggested that the ones they found inspired them to think of \consequences more broadly. The finding also suggests that \name could be a useful resource for gathering \consequences to include in a paper's ethics statement, well beyond the average of 0.6 words that are included in ethics statements in  NeurIPS AI papers, for example~\cite{ashurst2022, nanayakkara2021unpacking}. 

 Our follow-up interview and survey, however, painted a more complicated picture of anticipating \consequences using \name in practice. After using \name for their own projects, two of six participants in the second study were neutral on \name's usefulness for their projects, though all agreed that the system is useful \emph{for others}. Some participants acknowledged that they are not in the right position to think about these issues, though they learned ``a lot about social impact.'' The result resonates with the narrative reflected in Do et al.'s work~\cite{Do2023ThatsIB} that CS researchers tend to deflect the responsibility to consider the adverse effects of technology. %

 This brings us to how we see \name can support researchers in the future. Participants suggested that they would use \name for inspiration when writing broader impact statements in papers, which could lower the perceived burden of writing them~\cite{Abuhamad2020LikeAR} and potentially counteract the focus on desirable outcomes~\cite{sveiby2009unintended}. However, ultimately it would be ideal if CS researchers routinely learn about, anticipate, and reflect on \consequences---as has been repeatedly advocated for~\cite{Hecht2021ItsTT, Matthews2022EmbracingCV}---and that they do so early and proactively when addressing \consequences is still feasible~\cite{Do2023ThatsIB}. We envision \name as a tool that supports researchers in doing so, both by appealing to their intrinsic curiosity and by having extrinsic incentives such as fulfilling the requirements of a conference, grant agency, and institution. 
 
 \rr{Per Study 2}, \name, while a useful tool, is not a magical bullet to achieve this paradigm alone. Doing so would not only require a catalog of concrete consequences by \name but also a systematic change in culture and structural incentives. Tools like \name may spark new conversations and alleviate the perceived burden of anticipating \consequences particularly if research institutions evolve to actively encourage this reflection. For example, researchers can easily explore a wide variety of \consequences in their domain for past incidents before launching a new project. When writing ethics statements (e.g., for papers or grant proposals), researchers may use \name to efficiently and thoroughly examine their case. They could engage with the information and stay updated on the latest \consequences. A crucial aspect of this institutional change is nurturing a mindset among researchers that recognizes the importance of contemplating these challenges, thereby embedding the practice of considering \consequences as a fundamental aspect of responsible research. 
 
 In the long run, we envision \name to become an integral part of the technology development and research process by using strong incentives and implementing systemic changes as suggested in prior work~\cite{Hecht2021ItsTT,bernstein2021esr}. We hope that using \name will inspire the research community to work towards such a future in which learning about and anticipating \consequences soon becomes the norm.

%% file: sections/09-future.tex
\vspace{-0.3cm}
\section{Limitation \& Future Work}

 A limitation of the \name prototype is that it currently only uses online technology magazines and CHI papers to retrieve \consequences of technology. These may not be a comprehensive catalog of \consequences. In particular, the catalog may not adequately reflect the consequences that diverse user groups experience given that these articles are commonly written for  `tech-savvy' audiences. 
 In future work, we plan to systematically explore the difference in the reporting of \consequences across tech magazines, newspapers, and research papers from diverse fields and augment \name's sources. Future work should also incorporate non-English language articles, or non-American media outlets, to better reflect the effect of technology on diverse users. Another improvement is to incorporate multiple aspects rather than one category.
 Encouraged by the feedback from our participants, we also believe that there are exciting opportunities to enable citizen scientists to document their personal experiences with \consequences in \name. This could satisfy users' desire to share their own experiences while enabling insights into potential differential effects of technology on people. 




 We designed \name using LLMs due to the increasing performance on NLP tasks such as classification and summarization~\cite{bubeck2023sparks}. Nevertheless, LLMs can also introduce serious \consequences, such as model hallucination, biases in the training data, and a limited understanding of emerging fields. Our work extracts relevant information directly from trusted sources (i.e., online articles and papers) and provides access to the original content, instead of directly prompting LLMs to generate the information. 
In this paper, we offered a preliminary evaluation of the individual components within \name, using quantitative metrics such as F-1 score. While our metrics are comparable to similar ML tasks (see Section \ref{sec:case_study}), there is a tradeoff between foraging consequences at scale and ensuring perfect accuracy.
Involving citizen scientists could help improve the overall quality of the \name data curation process, such as by providing their feedback on the article relevance and label accuracy and by creating their summary and labels on the existing articles.


 We also foresee several different use cases for \name. For example, our participants indicated that \name could help researchers seek inspiration when writing broader impact statements for publications or grant proposals. Researchers may use \name to introduce the background knowledge in their areas to highlight their solutions to the public.  
 Additionally, policymakers and practitioners could use it to inform their work. Journalists could leverage a system like \name to discover new angles when reporting news, especially technology mishaps (see a related tool specifically designed for journalists for story inspiration~\cite{Maiden2018MakingTN}). The interested public can fairly efficiently learn about how digital technology has already affected, and will continue to affect, their lives. We believe that \name could also aid different stakeholders in reflecting on societal issues. 